\documentclass[prb,twocolumn,showpacs,floatfix,amsmath,amssymb]{revtex4-1}
\usepackage{graphicx}
\usepackage{amsmath,amssymb}
\usepackage{color}
\usepackage[T1]{fontenc}

\def\bra#1{\left\langle #1 \right|}             
\def\ket#1{\left| #1 \right\rangle}             
\def\tr{{\rm tr}}
\def\eref#1{\eqref{#1}}
\def\Or#1{\mathcal{O}\left({#1}\right)}
\def\fref#1{figure~\ref{#1}}
\def\Fref#1{Figure~\ref{#1}}
\def\beq{\begin{equation}}
\def\eeq{\end{equation}}
\def\beqn{\begin{eqnarray}}
\def\eeqn{\end{eqnarray}}

\def\ket#1{\vert #1 \rangle}

\renewcommand{\bf}{\mathbf}

\def\tr{\mathrm{tr}}
\begin{document}

\title{Quantum interference and Aharonov-Bohm oscillations in topological insulators}
\author{Jens H. Bardarson}
\author{Joel E. Moore}
\address{Department of Physics, University of California, Berkeley, CA 94720}
\address{Materials Sciences Division, Lawrence Berkeley National Laboratory, Berkeley, CA 94720}

\begin{abstract}

Topological insulators have an insulating bulk but a metallic surface.  In the simplest case, the surface electronic structure of a 3D topological insulator is described by a single 2D Dirac cone. A single 2D Dirac fermion cannot be realized in an isolated 2D system with time-reversal symmetry, but rather owes its existence to the topological properties of the 3D bulk wavefunctions.  The transport properties of such a surface state are of considerable current interest; they have some similarities with graphene, which also realizes Dirac fermions, but have several unique features in their response to magnetic fields.  In this review we give an overview of some of the main quantum transport properties of topological insulator surfaces. We focus on the efforts to use quantum interference phenomena, such as weak anti-localization and the Aharonov-Bohm effect, to verify in a transport experiment the Dirac nature of the surface state and its defining properties. In addition to explaining the basic ideas and predictions of the theory, we provide a survey of recent experimental work.
\end{abstract}

\maketitle
\tableofcontents

\section{Introduction}

Topological insulators (TI) are bulk insulators with protected metallic surface states as a result of the topological properties of the bulk electronic wavefunctions.\cite{Hasan:2010ku,Moore:2010ig,Qi:2011hb}  For a three-dimensional topological insulator,\cite{Hasan:2011hs} this metallic state is a  two-dimensional electron gas with many special features such as spin-momentum locking and a robustness to localization by disorder.  While several of these features have been observed with surface sensitive probes such as angle-resolved photoemission spectroscopy, much of current experimental focus is aimed at demonstrating these and other surface state properties in transport measurements.

The main goal of this review article is to explain how the Aharonov-Bohm and other magnetotransport effects are manifested in topological insulator surface states and to summarize recent experimental and theoretical progress towards their observation. The Aharonov-Bohm effect can be utilized as a fundamental probe of how the quantum phase of an electronic wavefunction is sensitive to magnetic flux through the gauge invariance of the Schr\"odinger equation coupled to electromagnetic fields.\cite{Aharonov:1959js,Aronov:1987ce,Batelaan:2009fm} It is possible to give self-contained explanations of how the features of the surface state, which is modeled by a single massless Dirac cone in the simplest case, lead to unusual (as compared to in a traditional two-dimensional electron gas) magnetotransport behavior in a variety of experimentally relevant situations.  This approach leaves out only the connection between bulk wavefunctions and surface electronic states, which requires a little bit of mathematical background and has been explained several times, for example in the reviews cited above.\cite{Hasan:2010ku,Moore:2010ig,Qi:2011hb,Hasan:2011hs}

In the remainder of the introduction we discuss the basic properties of topological insulator surface states, concentrating on what makes them different from the two-dimensional electron gas (2DEG) in either conventional semiconductor heterojunctions\cite{Ando:1982jy} or graphene.\cite{CastroNeto:2009cl,DasSarma:2011br} The main properties of the most common topological insulator materials are summarized and the relevant elements of quantum transport including the Aharonov-Bohm effect\cite{Aharonov:1959js,Aronov:1987ce} and localization theory\cite{Bergmann:1984bf,Lee:1985hn,Evers:2008gi} is given. While this review is mainly focused on the transport properties of 3D topological insulators, we comment at various points on the two-dimensional topological insulator or quantum spin Hall state,\cite{Konig:2008bz} as observed first in (Hg,Cd)Te quantum wells.\cite{Konig:2007hs} This phase is is interesting in itself and also illuminates some aspects of the three-dimensional behavior.

The main text consists of four sections. In section~\ref{sec:cond} we discuss the longitudinal conductivity which in the ideal case of insulating bulk should show ambipolar Hall effect and minimal conductivity when the chemical potential is tuned through the Dirac point. The main theoretical result covered is the absence of localization and the accompanying flow to the symplectic metal phase. Section~\ref{sec:field} addresses magnetic field induced quantum oscillations such as the Shubnikov-de Haas (SdH) oscillations. We focus in particular on the signatures of the Berry phase of the Dirac fermion in the SdH signal. The quantum Hall effect is briefly discussed. Magnetic flux effects on quantum transport, such as weak anti-localization (WAL) and Aharonov-Bohm (AB) oscillations, are the subject of section~\ref{sec:flux}. WAL is a useful probe of 2D transport but is insensitive to the Berry phase. In contrast, the AB oscillations can in principle be used to infer the presence of a nontrivial Berry phase. Finally, in section~\ref{sec:related} we collect some related problems and future directions. We end the review with a summary and conclusion.

\subsection{Surface states of three-dimensional topological insulators}

\begin{figure}[tb]
	\begin{center}
		\includegraphics[width=0.9\columnwidth]{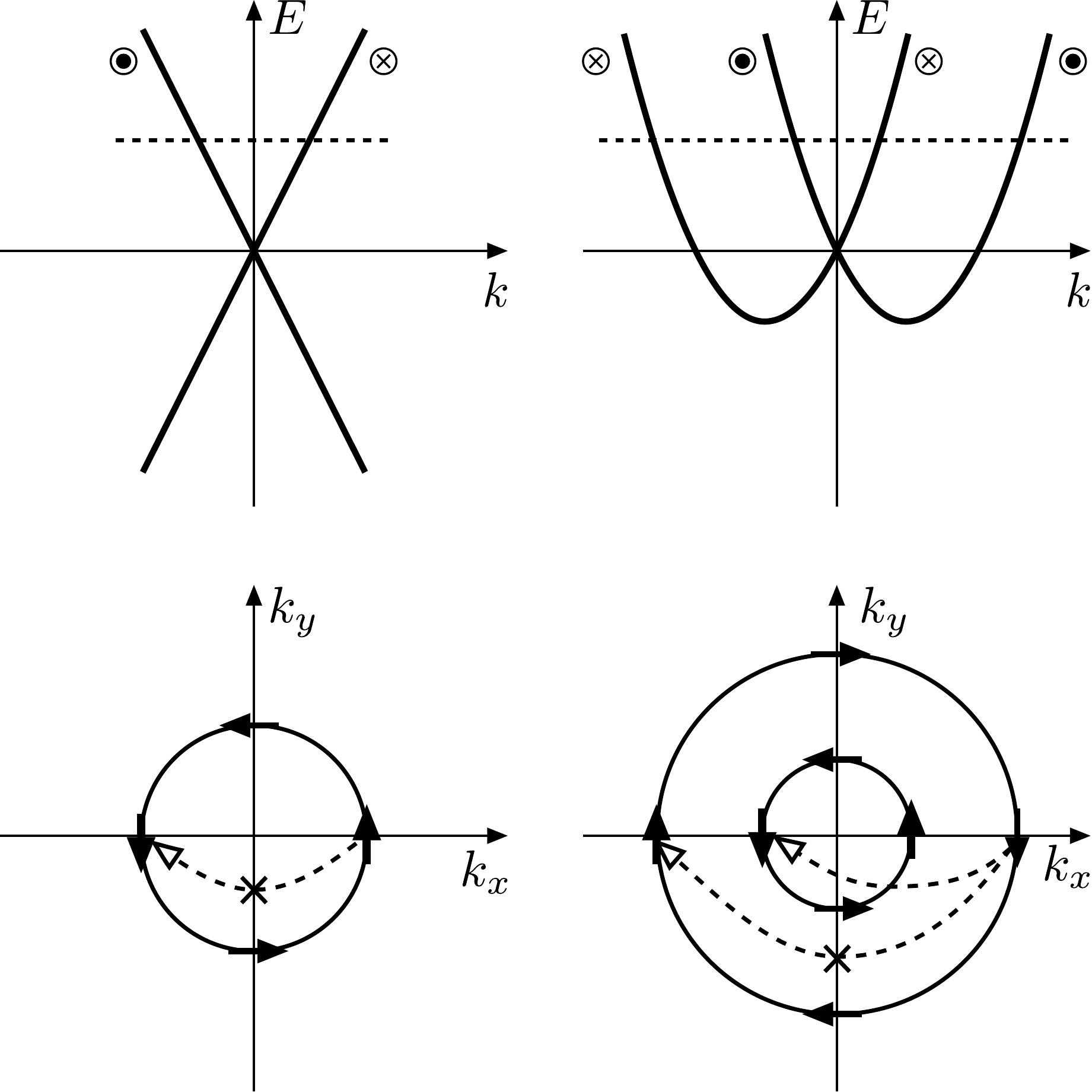}
	\end{center}
	\caption{Comparison between the Dirac dispersion (left top and bottom) and the dispersion of a 2DEG with Rashba spin-orbit coupling (right top and bottom). Both dispersions are rotationally symmetric around the energy axis, as shown by the plot of the Fermi surface in the lower panel. The Fermi surface consists of an odd and even number of closed curves (sheets) respectively. The filled and crossed circles and arrows denote the spin direction of the corresponding eigenstate. Due to time-reversal symmetry, scattering from a spin up state at $k$ to a spin down state at $-k$ is forbidden (crossed out dashed arrowed lines). Backscattering is therefore completely absent for the Dirac dispersion. In the Rashba case, scattering between the branches with the same spin state is allowed (dashed line without a cross), and backscattering can take place.}
	\label{fig:dispersions}
\end{figure}

In this section we introduce some key definitions and explain how the surface state of a topological insulator differs from a conventional two-dimensional electron system (cf.~\fref{fig:dispersions}).  Any strictly two-dimensional metal with time-reversal symmetry has a Fermi surface that consists of an even number of ``sheets'' (closed curves) once spin is included.\cite{Nielsen:1981ea,Nielsen:1981kj}  In the simple case of no spin-orbit coupling, the two sheets are degenerate; they move apart when spin-orbit coupling is included, but at every value of the Fermi energy the Fermi surface still consists of an even number of closed curves.  The surface state of a topological insulator is not strictly two-dimensional in the sense that it consists of a boundary between two different three-dimensional bulk states, one of which is frequently the vacuum. It has an odd number of Fermi surface sheets, and in the simplest case that odd number is 1.

As an example, consider the model linear dispersion relation obtained with the Hamiltonian 
\beq
H = v (p_x \sigma_y - p_y \sigma_x),
\label{eq:diracdispersion}
\eeq
where $v$ is the Fermi velocity, $\mathbf{p}$ the momentum, and $\sigma$ are the Pauli matrices. Another commonly used Hamiltonian $H = v \mathbf{p}\cdot\mathbf{\sigma}$ differs only in the interpretation of the sigma matrices and their relation to the real spin. For experimentally relevant materials the details can be more complicated: the velocity is not simply scalar but depends on the direction,\cite{Fu:2009ey,Alpichshev:2010kg,Kuroda:2010ha} the spin polarization may be reduced from its maximal value,\cite{Yazyev:2010jo,Pan:2011be,Souma:2011ci,Okamoto:2012wq} there may be more sheets of the Fermi surface\cite{Hsieh:2008ie} (i.e., an odd number larger than 1) and nonlinearity of the spectrum shows up away from the Dirac point.\cite{Shan:2010iy,Culcer:2010kf} Details of the surface geometry can also influence the electronic and spin structure.\cite{Silvestrov:2012hh,Zhang:2012cn,Zhang:2012uw} However, the Aharonov-Bohm effect and other magnetotransport phenomena discussed below are generally independent of these details.  The ``Dirac cone'' energy-momentum relation described by~\eref{eq:diracdispersion} is shown in~\fref{fig:dispersions}. 

Time-reversal symmetry implies that the state at momentum ${\bf p}$ must have spin direction opposite to that at $-{\bf p}$. The spin direction therefore precesses as the electron momentum moves around the Fermi surface. This is often referred to as spin-momentum locking. This case is different from a ``half-metal'' where the spin polarization is constant and time-reversal symmetry is broken.  As another comparison, adding the quadratic term $p^2 / 2m$ to the above Hamiltonian gives a Hamiltonian commonly used to describe quantum wells with Rashba spin-orbit coupling.\cite{Winkler:2003wv}  At every energy, there are either zero or two sheets of the Fermi surface (see~\fref{fig:dispersions}).

In 3D, the bulk wavefunctions of a perfect crystal are characterized by four topological invariants that take values in $\mathbb{Z}_2$, three ``weak'' and one ``strong''.\cite{Moore:2007gq,Fu:2007io,Roy:2009kk} The latter is the one of greatest interest both experimentally and theoretically, and we will use the term ``topological insulator'' to refer to strong topological insulators where this invariant is nonzero. The existence of an odd number of Fermi surface sheets is a consequence of the odd value of the strong topological invariant. There are also some interesting features if the strong index is zero but one of the weak invariants is nonzero.  Such ``weak topological insulators'' (WTI) have an even number of Dirac cones on the surface and can be viewed as layered versions of the two-dimensional topological insulator or quantum spin Hall state. The surface states of WTI's are in principle less stable to either bulk or surface disorder since there exists perturbations that gap out the surface. It turns out, however, that when this perturbation is zero on average as in a random environment, the surface state is robust against localization. We discuss this in more detail in section~\ref{sec:WTI}.

The low energy electronic structure of graphene is also described by two Dirac cones (ignoring the spin which does not play an important role in most graphene experiments). Graphene, however, has a time-reversal $\mathcal{T}$ with $\mathcal{T}^2 = +1$ and is therefore in a different symmetry class from TI's which time-reversal symmetry satisfies $\mathcal{T}^2 = -1$. In the absence of intervalley coupling an effective time-reversal symmetry with $\mathcal{T}^2 = -1$ emerges in graphene,\cite{Suzuura:2002hr} and the physics is that of a single Dirac cone. It is useful to keep this in mind since several results originally obtained for graphene are relevant to TI transport. 

The two-dimensional topological insulator or quantum spin Hall state has the same locking of spin and momentum in its edge state as in the surface state described in~\eref{eq:diracdispersion}: electrons moving one way along the edge have a certain spin orientation which is opposite that of the spin of the electrons moving in the reverse direction. There is a single branch of edge excitations moving in each direction, unlike in an ordinary quantum wire, which has two branches (spin-up and spin-down).

One simple difference in surface state transport between the 2D and 3D topological insulators can now be explained, and this will also help convey the importance of time-reversal symmetry.  Time-reversal symmetry implies that every spin-half eigenstate is degenerate in energy with, and distinct from, its time-reversal conjugate (the state obtained by reversing the direction of time).  As a result, every energy eigenvalue in a time-reversal invariant system of independent electrons is at least two-fold degenerate; these degenerate pairs are called Kramers pairs.  Integer-spin particles can be equivalent to their time-reversal conjugates and there need not be such degeneracies.  Now consider perturbing the original Hamiltonian.  The robustness of Kramers pairs necessitates that any time-reversal invariant perturbation, such as potential scattering, has a zero matrix element between the two states of a pair, as otherwise it would split the pair.

As a consequence of the robustness of the Kramers pairs, elastic scattering at the edge of a 2D topological insulator disappears at low energy (in the gap), because the two available states belong to the same Kramers pair.  The corresponding scattering from spin-up to spin-down is still forbidden in the case of an ordinary wire, but the scattering process that does not flip the spin is allowed and eventually leads to localization by disorder.  In general, an even number of Kramers pairs of edge modes will localize, while an odd number will lose pairs until a single pair is left which cannot be localized.  At low voltage and temperature, transport is effectively ballistic because backscattering disappears, although the corrections to the quantized conductance $e^2/h$ are expected to be power-law rather than exponential as in the quantum Hall case.\cite{Schmidt:2012iw}

The same Kramers protection exists at the surface of a 3D topological insulator but is less powerful as now there are allowed scattering processes that do not violate the Kramers theorem.  Scattering at any angle other than 180 degrees is allowed, and indeed Fourier transforms of STM measurements\cite{Roushan:2009kk} show the vanishing amplitude of perfect backscattering.  Because there are still allowed scattering processes at leading order, unlike in the 2D case, the low-temperature fate of conduction at the surface of 3D topological insulator requires more thought and is discussed below in the context of weak localization theory.

\subsection{Properties of topological insulator materials}

In this review, we will focus on the basic physics of the topological insulator phase revealed through Aharonov-Bohm measurements and other magnetic effects on transport.  However, in order to understand experiments, it seems useful to provide a few notes on the materials studied in current experiments, even though material improvements are rapid.  In two dimensions the first demonstration  of the theoretically predicted ``helical'' edge state was in (Hg,Cd)Te quantum wells.\cite{Konig:2007hs}  Recently, experiments were reported showing evidence for helical edge channels in InAs/GaSb quantum wells.\cite{Knez:2011hm}
There are theoretical proposals to realize the phase in other systems, e.g., when heavy atoms are adsorbed on graphene to increase the spin-orbit coupling\cite{Weeks:2011cu,Hu:2012tq} and in strained graphene in the presence of interactions.\cite{Abanin:2012iv}

In the remainder of this section, we concentrate on the 3D state, where there are more materials and experiments.  Bi-Sb alloys were the first materials studied for topological insulator behavior,\cite{Hsieh:2008ie} but the high level of alloy disorder and the complicated surface Fermi surface (with 5 band crossings along the cut studied) have led to their being superseded by other materials, in particular the semiconductors Bi$_2$Se$_3$, Bi$_2$Te$_3$, Bi$_2$Te$_2$Se, and variations thereof.\cite{Zhang:2009ks,Valla:2012ua}

Bi$_2$Se$_3$ is the 3D TI material that has been most investigated experimentally.\cite{Xia:2009ir}  It has a trigonal unit cell with five atoms, and can be pictured as having five layers Se-Bi-Se-Bi-Se; the central Se layer is clearly inequivalent to the outer two layers, and there is a good cleave plane between the van der Waals-bonded first and last Se layers.  The bulk bandgap is approximately 0.3 eV.  The surface Dirac velocity depends somewhat on the energy and the direction (there is a significant hexagonal distortion except very close to the Dirac point,\cite{Kuroda:2010ha}) but an approximate value useful for theoretical estimates is $4 \times 10^5$ m/s.  These numbers are consistent with estimates from GW-improved DFT calculations.\cite{Yazyev:2012ia}  Estimates of the surface state spin polarization from photoemission and numerical calculations range from 60\% of the spin half maximum up to nearly 100\%.\cite{Yazyev:2010jo,Pan:2011be}

Bi$_2$Te$_3$, also shown to be a topological insulator,\cite{Chen:2009do} has been studied for many years as a practical room-temperature thermoelectric material.  It has the same structure as Bi$_2$Se$_3$ and a smaller bulk bandgap of 0.15 eV.   Its thermoelectric utility can be understood from the rule of thumb that the ideal operating temperature of a thermoelectric semiconductor is about one-fifth of the bandgap $k_B T \approx E_g / 5$. A significant power factor (product of electrical conductivity, temperature, and thermopower squared) depends on having an appreciable number of thermally excited carriers.  Bi$_2$Te$_2$Se (``BTS'') has been studied very actively because crystals can be grown with bulk conductivity orders of magnitude lower than either Bi$_2$Se$_3$ or Bi$_2$Te$_3$.\cite{Jia:2011cn,Xiong:2012ht}  The Se atoms go in the middle layer of the five-layer structure. As opposed to Bi$_2$Se$_3$, which has the Dirac point in the bulk gap, the Dirac point in Bi$_2$Te$_3$ is buried deep in the valence band. 

Considerable effort is going into finding 3D topological insulators and related phases in other materials families, whether to decrease the bulk conductivity or to find ordered phases that combine topological order with another type of order (e.g., antiferromagnetism\cite{Mong:2010ka} or superconductivity.\cite{Fu:2010js}) The magnetotransport effects described below may be useful in identifying materials in the topological insulator phase when angle-resolved photoemission or other methods are impractical. In addition to the Bi based materials, straining 3D HgTe which is nominally a semimetal, opens up a gap and realizes a TI.\cite{Brune:2011hi} $\beta$-Ag$_2$Te has also been used in transport studies.\cite{Zhang:2011du,Sulaev:2012wa} 

In early transport experiments the conductance was dominated by the bulk. The cleanest signatures of quantum interference from the surface state have been obtained in thin films and nanowires. These are either epitaxially grown or obtained with mechanical exfoliation. Due to the easy cleave plane, the thickness is generally a multiple of quintuple layers with each quintuple layer about 1 nm thick. In the ultrathin limit the tunnel coupling between the top and bottom surface is sufficiently large to open up a sizable gap and make the film insulating.\cite{Linder:2009ks,Liu:2010fo,Lu:2010fr} This gap is observed experimentally in films a few nanometers thick\cite{Zhang:2009jc,Liu:2012hf} and its rapid decay to zero with increasing thickness gives a direct measurement of the surface state's penetration into the bulk that can be compared with photoemission results.\cite{Zhang:2010br} Insulating behavior in transport has also been observed.\cite{Cho:2011ht,Cho:2012de,Taskin:2012kx}

With increased surface mobility the possibility of realizing correlation physics opens up. One direction is fractional quantum Hall physics at the surface; indeed some features are observed in transport data that are conjectured as possibly indicating incipient fractional Hall states at the surface.\cite{Analytis:2010kl} A thin film of topological insulator in a magnetic field is analogous to a quantum Hall bilayer.\cite{Eisenstein:2004go} These bilayers have been a fertile system for studies of many-body physics, as the proximity of the two layers enables interlayer correlated phases with remarkable properties when the Coulomb interaction between the layers is significant.  However, these correlation effects may be hidden by the hybridization gap in topological insulator thin films.

\subsection{Diffusive transport and the symplectic metal}
\label{sec:SymplecticMetal}
\begin{figure}[tb]
	\begin{center}
		\includegraphics[width=0.9\columnwidth]{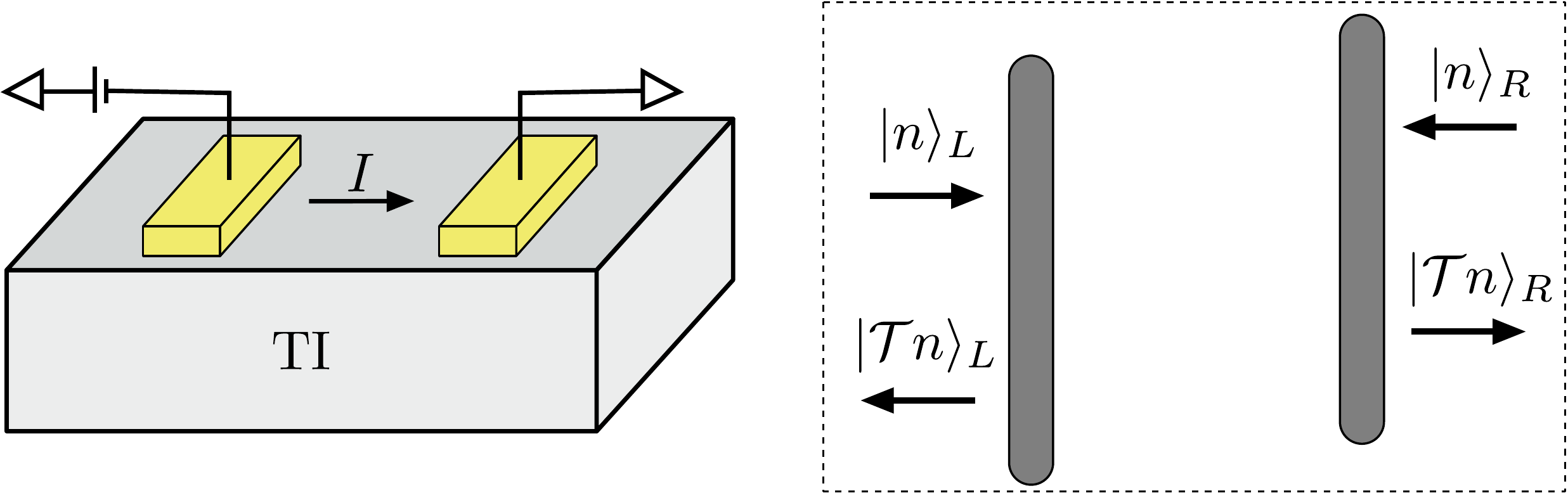}
	\end{center}
	\caption{A schematic of a typical two terminal transport setup (left), and the ideal modeling with a set of incoming and outgoing modes (right). A current is injected through one lead and extracted through the other. In the ideal model, the current is assumed to go only through the surface and not through the bulk. We allow for the possibility that the modes are different in the left (L) and right (R) lead, though in practice the modes in the two leads are often related to each other. For example, in the Dirac case~\eref{eq:DiracModes} $\ket{n}_L = \ket{\mathcal{T}(-n)}_R$ up to a possible phase.}
	\label{fig:TwoTerminalTransport}
\end{figure}

In the Altland-Zirnbauer symmetry classification,\cite{Altland:1997cc} the topological insulators we are interested in here belong to the symplectic class (AII in the Cartan notation). This symmetry class is characterized by the presence of a time-reversal symmetry. The time-reversal operator $\mathcal{T}$ commutes with the Hamiltonian and satisfies $\mathcal{T}^2 = -1$. A two dimensional electron gas (2DEG) with strong spin-orbit coupling is also in the symplectic class, as well as graphene in the absence of intervalley scattering.\cite{Suzuura:2002hr}
When the electronic transport is diffusive many of its characteristics are independent of the topological properties of the underlying system. We will refer to this state as the {\it symplectic metal}.  The symplectic metal is characterized by weak anti-localization (WAL). Because of WAL the symplectic metal is, in the renormalization group sense, a stable fixed point. In this section we review the various characteristic quantum interference phenomena of the symplectic metal. We limit the discussion to two terminal transport. This ideal setup can for example model transport in one side of a large TI surface or conductance through a TI nanowire as schematically shown in~\fref{fig:TwoTerminalTransport}.

The presence of a time-reversal symmetry crucially affects many of the transport properties of the symplectic metal. Before discussing the perhaps more intuitive path picture of the interference phenomena, we consider how these directly result from the symmetry constraints imposed by time-reversal symmetry on the scattering matrix describing the two terminal transport. In particular, we obtain absence of backscattering,\cite{Ando:1998fn} discuss under what condition a perfectly transmitted mode is realized,\cite{Ando:2002es} and give the connection between the time-reversal symmetry and weak anti-localization. 

\subsubsection{Quantum transport and time-reversal symmetry}
The scattering matrix describing the two terminal transport in the presence of time-reversal symmetry with $\mathcal{T}^2=-1$ can be chosen to be antisymmetric. Before discussing the consequences of this antisymmetry we give a short derivation of this fact, following Ref.~\onlinecite{Bardarson:2008jk}. 

Consider a two terminal setup as in~\fref{fig:TwoTerminalTransport}. The left and right leads are metallic contacts which host a large number of incoming modes, denoted by $\ket{n}_L$ and $\ket{n}_R$ respectively. In a topological insulator these modes are the properly normalized eigenstates of the Dirac Hamiltonian~\eref{eq:diracdispersion}. For example, if we assume periodic boundary conditions in the transverse $y$ direction, the modes can be written
\begin{eqnarray}
	\ket{n}_L &= \frac{1}{\sqrt{2}}
	\begin{pmatrix}
		1 \\ i
	\end{pmatrix}e^{ikx+iq_ny},
	\\
	\ket{n}_R &= \frac{1}{\sqrt{2}}
	\begin{pmatrix}
		1 \\ -i
	\end{pmatrix} e^{-ikx+iq_ny}.
	\label{eq:DiracModes}
\end{eqnarray}
Here $q_n = 2\pi n/W$ are the discrete transverse momenta with $W$ the transverse width of the sample, and the Fermi energy $E_F = \hbar v\sqrt{k^2+q_n^2}$. We have assumed $k \gg q_n$ such that we can ignore the momentum dependence of the spinors. This is the relevant limit in the leads, which being metallic are highly doped.\cite{Tworzydio:2006hw} These modes carry unit current $\bra{n}\sigma_x\ket{n}_L = -\bra{n}\sigma_x\ket{n}_R = 1$.
The outgoing modes are the time-reverse of the incoming modes $\ket{\mathcal{T}n}_L$ and $\ket{\mathcal{T}n}_R$, since time-reversal reverses the direction of motion. Importantly, $\ket{\mathcal{T}n}$ has the opposite spin to $\ket{n}$. 

A solution to the Hamiltonian describing the system, subject to the boundary conditions imposed by the metallic leads at $x = 0$ and $x=L$, can be written
\begin{eqnarray}
	\ket{\psi} =
	\begin{cases}
		\sum_n c_{n,L} \ket{n}_L + d_{n,L} \ket{\mathcal{T}n}_L,  & x \leq 0, \\
		\sum_n c_{n,R} \ket{n}_R + d_{n,R} \ket{\mathcal{T}n}_R,  & x \geq L, \\
		\ket{\psi_M},   & 0 \leq x \leq L.
	\end{cases}
	\label{eq:scatteringstate}
\end{eqnarray}
Since $\ket{\psi}$ is a solution to the Schr{\"o}dinger equation, the incoming coefficients $c_L$ and $c_R$ are linearly related to the outcoming coefficients $d_L$ and $d_R$ through the scattering matrix $S$
\begin{equation}
	\begin{pmatrix}
	 d_L \\
	 d_R
	\end{pmatrix}
	 = S
	 \begin{pmatrix}
	 c_L \\
	 c_R
 \end{pmatrix}
	\label{eq:Sdefinition}
\end{equation}
Assuming an equal number of modes $N$ on the left and the right, the $2N\times2N$ scattering matrix has the block diagonal form 
\begin{equation}
	S = \begin{pmatrix}
	r & t^\prime \\
	t & r^\prime
	\end{pmatrix}
	\label{eq:S}
\end{equation}
where $r_{nm}$ is the probability amplitude of reflection of mode $\ket{m}_L$ into $\ket{\mathcal{T}n}_L$, and the other block matrices have a similar interpretation. For our current purpose we do not need to know the exact form of the wavefunction $\ket{\psi_M}$ in the sample. In an actual calculation the scattering matrix is obtained from a knowledge of $\ket{\psi_M}$. The scattering matrix gives the two terminal conductance through the Landauer formula 
\begin{equation}
	G = \frac{e^2}{h}\tr\, t^\dagger t = \frac{e^2}{h}\tr\, (1-r^\dagger r), 
	\label{eq:Landauer}
\end{equation}
where the second equation follows from current conservation $S^\dagger S = 1$. For ease of notation we also introduce the unitless conductance $g = G/(e^2/h)$. The conductivity of a sample of width $W$ and length $L$ is given by $\sigma = G L/W$. 

Because of time-reversal symmetry, if $\ket{\psi}$ is an eigenstate of the Hamiltonian so is $\ket{\mathcal{T}\psi}$. This solution, furthermore, is orthogonal to $\ket{\psi}$ since
\begin{equation}
	\bra{\mathcal{T}\psi}\psi\rangle = \bra{\mathcal{T}^2\psi}\mathcal{T}\psi\rangle^* = -\bra{\mathcal{T}\psi}{\psi}\rangle = 0.
	\label{eq:KramersDegE}
\end{equation}
In the first equality we used the antiunitarity of $\mathcal{T}$, then that $\mathcal{T}^2 = -1$. $\ket{\mathcal{T}\psi}$ is therefore an independent solution and all energy eigenvalues are doubly degenerate: the celebrated Kramers degeneracy. As a consequence 
\begin{eqnarray}
	\ket{\mathcal{T}\psi} =
	\begin{cases}
		\sum_n c^*_{n,L} \ket{\mathcal{T}n}_L - d^*_{n,L} \ket{n}_L,  & x \leq 0, \\
		\sum_n c^*_{n,R} \ket{\mathcal{T}n}_R - d^*_{n,R} \ket{n}_R,  & x \geq L, \\
		\ket{\mathcal{T}\psi_M},   & 0 \leq x \leq L,
	\end{cases}
	\label{eq:Tscatteringstate}
\end{eqnarray}
is also an allowed scattering state. By inspecting~\eref{eq:scatteringstate} we obtain
\begin{equation}
	\begin{pmatrix} 
	c_L^* \\ c_R^* 
	\end{pmatrix} = -S 
	\begin{pmatrix}
	d_L^* \\ d_R^*
	\end{pmatrix}.
	\label{eq:TS}
\end{equation}
Comparing with Eq.~\eref{eq:Sdefinition} one concludes that time-reversal symmetry requires
\begin{equation}
	S^T = -S,
	\label{eq:Santisymmetric}
\end{equation}
with $T$ denoting the transpose. The antisymmetry of the scattering matrix gives rise to the absence of backscattering, the presence of a perfectly transmitted mode, and weak anti-localization, as we will now explain.

The absence of backscattering is simply the fact that the diagonal elements $r_{nn}=r^\prime_{nn} = 0$. This statement holds true both in a TI and a 2DEG, since it only requires the presence of a time-reversal symmetry. However, since $r_{nn}$ is reflection of mode $\ket{n}$ back into $\ket{\mathcal{T}n}$, which has the opposite spin, backscattering that does not flip the spin is allowed in the 2DEG (cf.~\fref{fig:dispersions}). There is no such state in a TI and therefore backscattering is completely absent.

Due to the antisymmetry of $r$, the eigenvalues of $r^\dagger r$, and by unitarity also the transmission eigenvalues of $t^\dagger t$, come in degenerate pairs. This is the Kramers degeneracy of transmission eigenvalues. Furthermore, if the number of modes $N$ is odd, there is necessarily at least one transmission eigenvalue that is equal to unity, since $\det(r) = \det(-r^T) = (-1)^N \det(r) = -\det r = 0$. This is the perfectly transmitted mode discussed by Ando and Suzuura,\cite{Ando:2002es} and it will play a central role in our discussion below. 

An odd number of modes can never strictly be realized in an inherently 2D system with time-reversal symmetry.\cite{Nielsen:1981ea,Nielsen:1981kj} The absence of certain couplings can, however, reduce the system to an effective one with an odd number of modes. An example of this is graphene in the absence of intervalley coupling.\cite{Suzuura:2002hr} In contrast, a TI can intrinsically have an odd number of modes and therefore host a perfectly transmitted mode.

Weak anti-localization is the first order in $1/g$ quantum correction to the classical Drude conductance. To obtain WAL from the scattering matrix we assume the elements of $S$ to be randomly distributed Gaussian variables\cite{Beenakker:1997gz} with
\begin{equation}
	\langle S_{nm}^*S_{nm} \rangle = \frac{1-\delta_{nm}}{2N-1},
	\label{eq:Srandom}
\end{equation}
where the angular brackets denote average over the distribution of the elements of the scattering matrix. The delta function is needed to satisfy the antisymmetry condition~\eref{eq:Santisymmetric} and the denominator is obtained from the unitarity of $S$. Using this in the Landauer formula~\eref{eq:Landauer}
\begin{equation}
	g =  N - \frac{N^2-N}{2N-1} = \frac{N}{2} + \frac{1}{4} + \Or{\frac{1}{N}}.
	\label{eq:WALrandomS}
\end{equation}
The first term is the classical conductance. It takes the value $N/2$ since each mode is equally likely to be transmitted as being reflected, due to the randomness of the scattering matrix. The second term is a positive enhancement of the conductance due to quantum interference, namely weak anti-localization. This term is absent in the absence of time-reversal symmetry, as is readily verified. A time-reversal breaking pertubation, such as a magnetic field, will therefore decrease the conductance. The fact that the first quantum correction is positive reflects the stability of the symplectic metal phase to weak disorder. This correction is perturbative (here in $1/N$) and independent of topology and is therefore the same for TI surfaces and a 2DEG. The case of strong disorder and small conductance is discussed in section~\ref{sec:AbsenceOfLoc}.

This argument is instructive in that it shows the relation between WAL and time-reversal symmetry. The assumption~\eref{eq:Srandom} is however only strictly valid when the two leads are connected by a quantum dot.\cite{Beenakker:1997gz} We are interested in the case of a two dimensional sample connecting the two leads. This will be discussed in the next section.

\subsubsection{Weak anti-localization and Berry phase}
When a Dirac fermion traverses a loop in space, the spin rotates by $2\pi$ due to the spin-momentum locking. The wave function, being a spinor, acquires a phase of $\pi$ which can alternatively be considered as a Berry phase induced by the Dirac point. This phase affects quantum interference and in particular changes the constructive interference of spinless fermions that gives rise to weak localization into destructive interference and WAL.

Studies of weak (anti)-localization date back a couple of decades and the physics is by now well understood. A number of reviews\cite{Bergmann:1984bf,Lee:1985hn} and textbooks (our discussion is of similar flavor as\cite{Akkermans:2007vf}) exist that discuss the basic phenomena. The importance of spin-orbit coupling and the resulting WAL correction was first derived by Hikami, Larkin and Nagaoka.\cite{Hikami:1980jn} An intuitive picture has been given by Bergmann.\cite{Bergmann:1983cy} The details of the derivation for Dirac fermions are slightly different, but the final answer for the WAL correction is the same, as shown by Suzuura and Ando\cite{Suzuura:2002hr} and McCann et al.\cite{McCann:2006ip} in the context of graphene. This is to be expected since the two systems are in the same universality class. Nevertheless, we include here a short introduction to the topic, focusing on the broad physical picture and avoiding detailed formalism. This should serve as a guide to the literature and to make this review more self-contained. In particular, we will need some of the results discussed in this section in our later survey of TI transport experiments. In the context of TI's the WAL correction has been discussed further theoretically by several authors.\cite{Imura:2009cm,Tkachov:2011em,Lu:2011iz,Lu:2011ja,Garate:2012fm,Adroguer:2012uc,Krueckl:2012ul}

\begin{figure}[tb]
	\begin{center}
		\includegraphics[width=0.9\columnwidth]{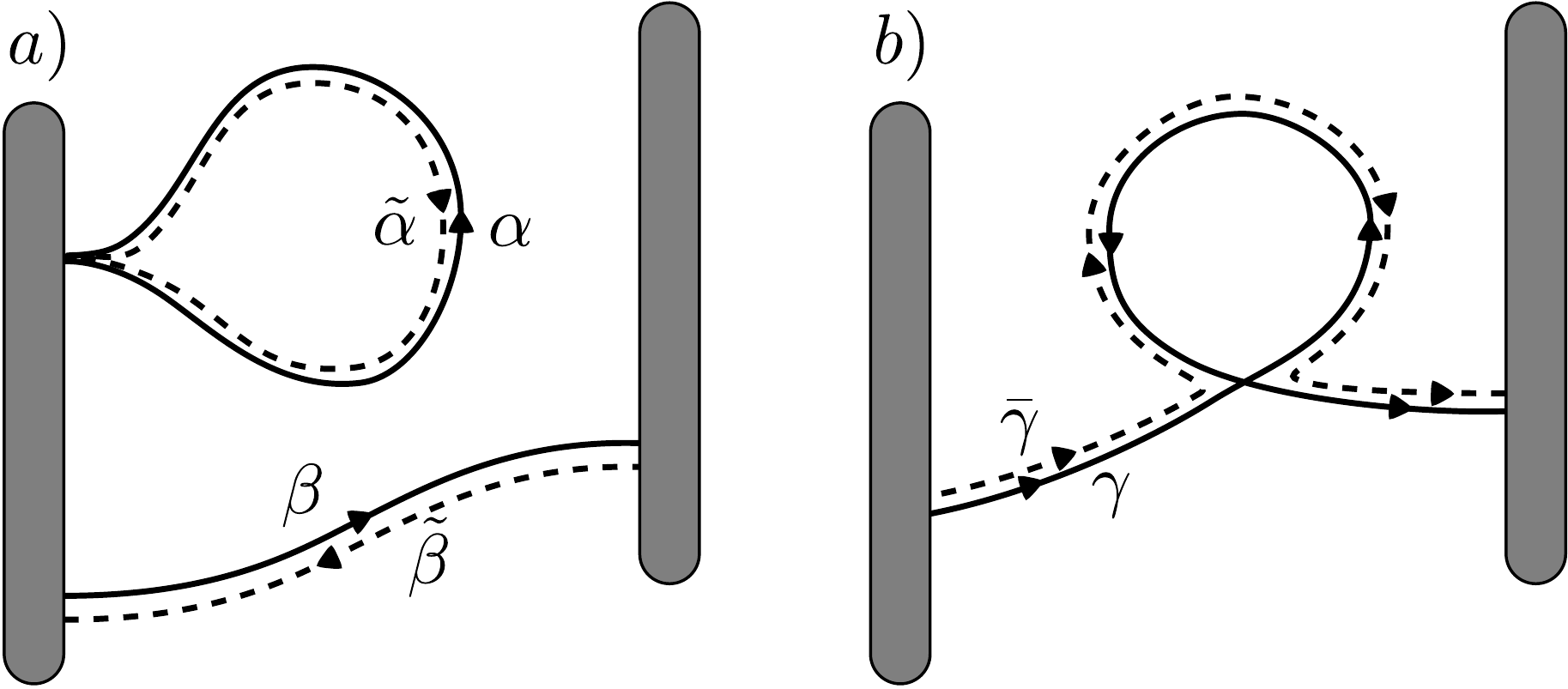}
	\end{center}
	\caption{Classical paths that interfere to give rise to weak anti-localization. a) The path $\alpha$ and its time-reverse $\tilde{\alpha}$, which return to the time-reversed mode in the lead, interfere destructively leading to the absence of backscattering. Paths $\beta$ going from one lead to the other do not interfere with their time-reversed partner $\tilde{\beta}$ since these paths start in the opposite lead. However, b) paths $\gamma$ ($\bar{\gamma}$) that have a single (avoided) crossing do interfere and including these paths is crucial to obtain a current conserving theory. In addition to the $\gamma$ paths shown, there will be similar paths that return back to the same lead after the crossing.}
	\label{fig:wal}
\end{figure}

As in the last section we focus on the two terminal setup and the conductance in terms of the scattering matrix. Reflection and transmission amplitudes can be written as a sum over all possible paths connecting the two leads (see~\fref{fig:wal}).
\begin{eqnarray}
	r_{nm} &= \sum_\alpha A_\alpha e^{iS_\alpha/\hbar}, \\
	t_{nm} &= \sum_\beta A_\beta e^{iS_\beta/\hbar}.
	\label{eq:rt_scl}
\end{eqnarray}
Here $\alpha$ denotes paths that enter the sample from mode $\ket{m}_L$ in the left lead and exit through mode $\ket{Tn}_L$ in the same lead. Similarly, $\beta$ denotes paths that start in mode $\ket{m}_L$ in the left lead and exit through mode $\ket{Tn}_R$ on the right. $S_\alpha$ is the classical action of path $\alpha$ and $A_\alpha$ contains the stability amplitude, normalization and possible geometric phases of the path $\alpha$.\cite{Gutzwiller:1990vo,Haake:2010tm} In particular, any phase acquired from rotation of the spin degree of freedom will enter through $A_\alpha$. Generally, the effect of the spin rotation is captured by a matrix valued amplitude\cite{Zaitsev:2005fe} but this is not needed for our current purpose due to the spin-momentum locking.

\begin{figure}[tb]
	\begin{center}
		\includegraphics[width=0.8\columnwidth]{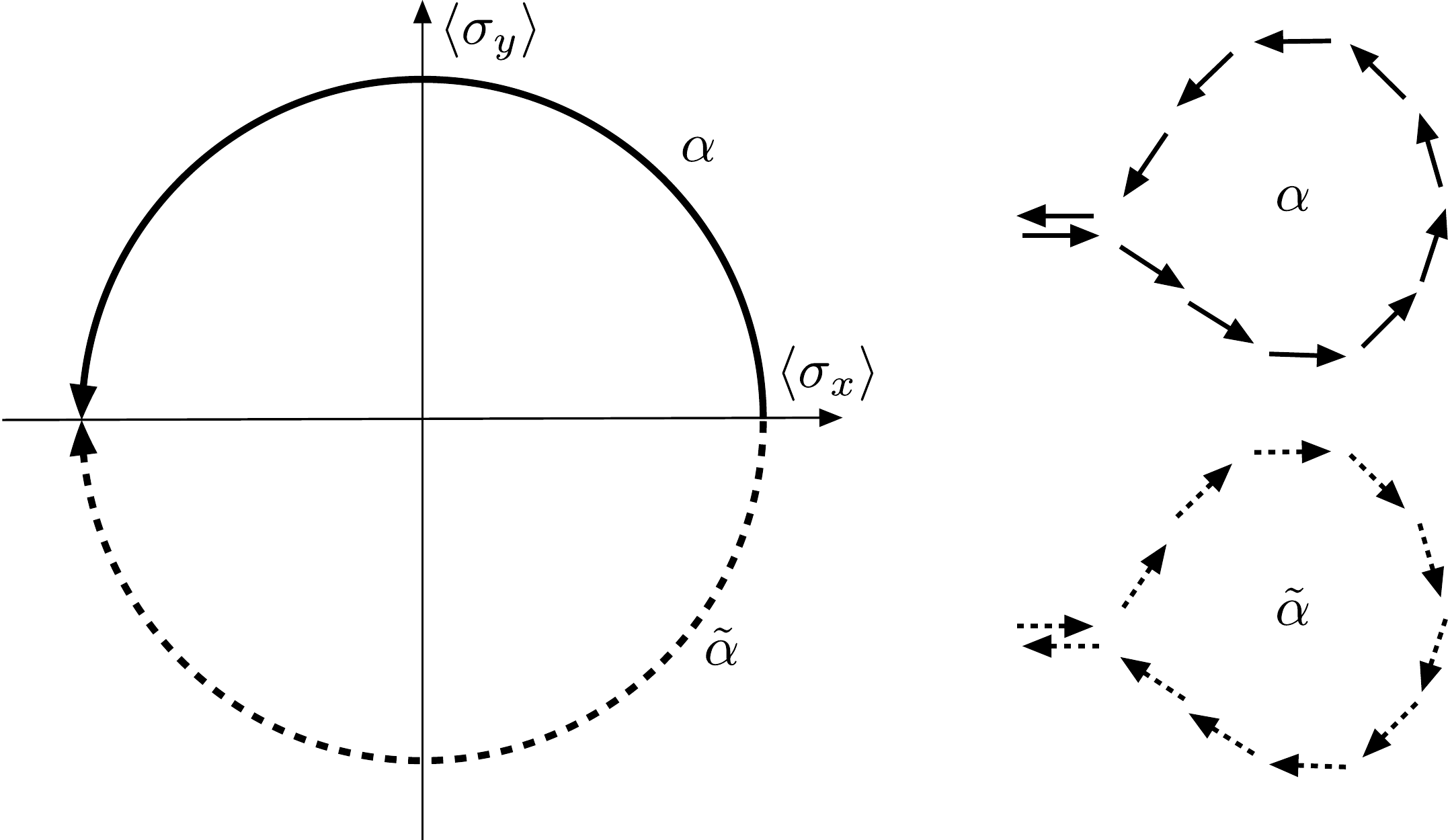}
	\end{center}
	\caption{The spin rotation along the paths $\alpha$ and $\tilde{\alpha}$. Since the paths enter with opposite sign in the interference term the total spin rotation is $2\pi$, resulting in the important minus sign that causes destructive interference in the reflection amplitude and the absence of backscattering.}
	\label{fig:BerrySpin}
\end{figure}

In terms of reflection amplitudes, the conductance~\eref{eq:Landauer} is given by
\begin{equation}
	g = N - \sum_{\alpha,\alpha^\prime}A_\alpha A_{\alpha^\prime}^* e^{i(S_\alpha-S_{\alpha^\prime})/\hbar}.
	\label{eq:Gr}
\end{equation}
When averaging over disorder, the quickly oscillating exponential term will generally average to zero, unless $S_\alpha \approx S_{\alpha^\prime}$. The classical Drude conductance is obtained by including only the diagonal terms $\alpha^\prime = \alpha$. In addition, interference terms where $\alpha^\prime = \tilde{\alpha}$ with $\tilde{\alpha}$ the time-reverse of $\alpha$ survive the disorder average since by time-reversal symmetry $S_\alpha = S_{\tilde{\alpha}}$. The amplitudes $A_\alpha$ and $A_{\tilde{\alpha}}$ have the same absolute value, but have a phase difference of $\pi$, and thus $A_\alpha = -A_{\tilde{\alpha}}$. This is due to the Berry phase picked up by the $2\pi$ relative rotation of the spin between the two paths, see~\fref{fig:BerrySpin}. Therefore the total contribution of these two paths to the sum in~\eref{eq:Gr} is 
\begin{equation}
	|A_\alpha|^2 + |A_{\tilde{\alpha}}|^2 + 2\Re (A_\alpha A^*_{\tilde{\alpha}}) = 0.
	\label{eq:backsc}
\end{equation}
This is the absence of backscattering $r_{nn} = 0$ obtained from symmetry in the last section. Compared to the classical value $2|A_\alpha|^2$, the probability to reflect back into the same lead is reduced and the conductance is enhanced. Note that this destructive interference only affects the diagonal elements of the matrix $r$ of reflection amplitudes.

This is not the full story, as one infers from the fact that such paths do not affect the conductance when written in terms of transmission amplitudes
\begin{equation}
	g = \sum_{\beta,\beta^\prime}A_\beta A_{\beta^\prime}^* e^{i(S_\beta-S_{\beta^\prime})/\hbar}.
	\label{eq:Gt}
\end{equation}
Indeed, if $\beta$ goes from the left lead to the right lead, the time-reversed path $\tilde{\beta}$ goes from the right lead to the left lead and therefore does not enter the sum~\eref{eq:Gt}. With no change to the transmission amplitudes as compared with the classical value, current is no longer conserved and the scattering matrix is not unitary if we include only these paths. To get a consistent, current conserving theory, we need to include additional paths. These are the paths $\gamma$ that have a single self crossing (\fref{fig:wal}), and the corresponding path $\bar{\gamma}$ which is almost the same except for an avoided crossing at the crossing point.\cite{Aleiner:1996ka} The two paths therefore traverse the loop in the path in opposite direction. The sign of the correction to the amplitude is determined by the phase picked up in the loop and the details of the encounter at the crossing.\cite{Richter:2002cn} Instead of going through the derivation, we can simply infer the sign of the correction from the absence of backscattering and the need to obtain a unitary scattering matrix. In addition to the paths $\gamma$ that end up in the opposite lead after leaving the loop, there are paths that instead return to the same lead and therefore affect reflection probabilities. The interference of the loop is independent of which lead the particle ends up in after leaving the loop. Thus, the correction from these loops has the same sign for reflection and transmission probabilities. This correction is therefore necessarily positive to compensate the negative contribution due to the absence of backscattering. Transmission is enhanced by the quantum interference giving rise to WAL. In this sense absence of backscattering and WAL are integrally related through current conservation. 

\begin{figure}[tb]
	\begin{center}
		\includegraphics[width=0.9\columnwidth]{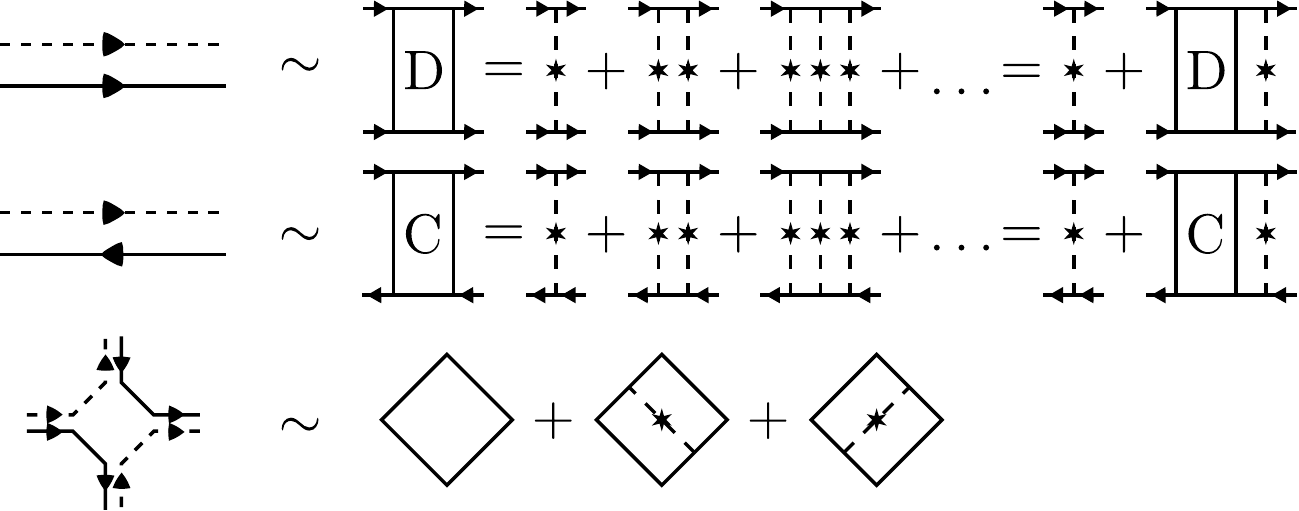}
	\end{center}
	\caption{The different parts of the interference between the paths with a single (avoided) crossing in \fref{fig:wal} can be interpreted as objects that are obtained in diagrammatic calculations. The initial and last part where the paths propagate together corresponds to the diffusion (top) while the propagation in the opposite direction in the loop corresponds to the cooperon (middle). The crossing where the paths switch between propagating in the same direction and opposite direction corresponds to the Hikami box in the diagrammatic language (bottom). Here we have shown the dressed Hikami box which is obtained since all the terms on the right are of the same order. The crossing on the left is topologically equivalent to crossing as plotted in~\fref{fig:wal}.}
	\label{fig:diagrammatics}
\end{figure}
In the next step we focus on the quantum corrections to the conductance as interference between the paths $\gamma$ and $\bar{\gamma}$ as depicted in the diagram of~\fref{fig:wal}b. Instead of having the diagram denote only two paths we reinterpret it as a sum of many paths. Decomposing the diagram into parts, each part has an interpretation which corresponds to an object that is commonly obtained in diagrammatic calculations of the conductivity. This connection between the path picture of WAL and the diagrammatic calculation aids the intuitive understanding of the latter.

To that end, we decompose the diagram of~\fref{fig:wal}b into three main parts.
The first part represents the sum of all possible pair of paths that travel together from the lead towards the crossing point. This includes paths that are scattered by a single impurity, two impurities, and so on. Since they travel together, the two paths scatter of the same set of impurities in the {\it same} order. This is equivalent to the ``diffusion'' in the diagrammatic language and naturally satisfies a diffusion equation. Essentially, a density disturbance is injected at the lead and diffuses towards the crossing point. The second part is the loop, which the two paths traverse in opposite direction. The paths again scatter of the same set of impurities, but now in the {\it opposite} order. In the diagrammatic language this corresponds to the ``cooperon''. When time-reversal symmetry is preserved, it also satisfies a diffusion equation. The last part is the crossing point, where the paths switch from traveling together to traveling in the opposite direction. This corresponds to the Hikami box in the diagrammatic language.

To find the correction to the conductivity we need to estimate the number of paths with a single self-crossing. This is equivalent to calculating the probability for the cooperon to return back to the crossing point. Since it satisfies a diffusion equation, in time $t$ this probability is proportional to $1/t$. It does not matter at what time the crossing happens, as long as the time is smaller than the phase coherence time $\tau_\phi$ after which the paths are no longer phase coherent. Integration over time gives  
\begin{align}
	\delta\sigma &= \frac{e^2}{2\pi h}\int_{\tau}^{\tau_\phi} \frac{1}{t} dt = \frac{e^2}{2\pi h}\ln \frac{\tau_\phi}{\tau} \notag \\
	& = \frac{e^2}{\pi h}\ln\frac{\ell_\phi}{\ell} = -p\frac{e^2}{2\pi h}\ln\frac{T}{T_\ell}.
	\label{eq:WAL}
\end{align}
The lower bound is given by the mean free time $\tau$, below which the motion is ballistic, and the dephasing length $\ell_\phi^2 = D\tau_\phi$ with $D$ the diffusion constant. In the last equality we have assumed the temperature dependence of the dephasing to be given by a power law $\tau_\phi \sim T^{-p}$ with $p>0$. $T_\ell$ is defined by the relation $\tau_\phi/\tau = (T/T_\ell)^{-p}$.  We have not explicitly kept track of the prefactors, since our main goal here is to give a phenomenological understanding of the origin of the logarithmic dependence of the WAL correction. Details can be found for example in Ref.~\onlinecite{Akkermans:2007vf}.

We have described a current conserving theory of WAL that takes into account paths with a single self crossing. In principle, there are additional paths with a larger number of crossings and these also give arise to interference correction. One can show that including only the single crossing paths is equivalent to assuming $k_F\ell \gg 1$, with $k_F$ the Fermi momentum and $\ell$ the mean free path.\cite{Akkermans:2007vf} This is the same condition required for the validity of most diagrammatic calculations. More crossings give higher order corrections in $1/k_F\ell$, and are eventually responsible for localization (see e.g.\ Ref.~\onlinecite{Muller:2007hb} and references therein for a description of such higher order paths in chaotic transport).

In theoretical work on quantum transport of Dirac fermions, the WAL correction~\eref{eq:WAL} is often used to identify the symplectic metal. For example, in section~\ref{sec:AbsenceOfLoc} on the absence of localization and section~\ref{sec:WTI} on transport in WTI's, a logarithmic dependence of the conductance on the system size (replacing the phase coherent length $\ell_\phi$ in finite systems) with a slope $e^2/\pi h$ is observed. Experimentally, is is often easier to observe WAL by applying a magnetic field. The magnetic field breaks time-reversal symmetry and changes the interference of the cooperon such that eventually the WAL correction disappears. This is the subject of the next section.

\subsubsection{Aharonov-Bohm and Altshuler-Aronov-Spivak magnetoconductance oscillations}
In the presence of a magnetic field perpendicular to the 2D motion, the paths $\alpha, \beta, \gamma$ pick up the Aharonov-Bohm (AB) phase\cite{Aharonov:1959js,Aronov:1987ce}
\begin{equation}
	A_\alpha \rightarrow A_\alpha e^{-i\frac{e}{\hbar}\int_\alpha \mathbf{A} \cdot d\mathbf{r}},
	\label{eq:AB}
\end{equation}
with $\mathbf{A}$ a vector potential of the magnetic field $\mathbf{B} = \mathbf{\nabla}\times\mathbf{A}$, and the integration is along the path $\alpha$. If the path is a closed loop, the integral is equal to the flux $\phi$ through the loop and the Aharonov-Bohm phase becomes $2\pi\phi/\phi_0$, with $\phi_0 = h/e$ the magnetic flux quantum. The AB phase can both lead to periodic oscillation of the conductance and destruction of WAL.

We assume the magnetic field is weak enough not to affect the classical motion on the time scale of the mean free time $\tau$, i.e.\ that $\omega_c\tau \ll 1$ with $\omega_c$ the cyclotron frequency. The diffusion constant, and consequently the Drude conductivity, therefore remains unchanged. In the diffusion the two paths pick up the same phase which is then canceled. In the loop (cooperon) the two paths pick up the opposite phase. This modifies the diffusion of the cooperon which now satisfies a diffusion equation with $\mathbf{p} \rightarrow \mathbf{p}-2e\mathbf{A}$. Solving the modified diffusion equation, one finds that the return probability is proportional to $B/\phi_0 \sinh (4\pi DtB/\phi_0)$.\cite{Akkermans:2007vf} The WAL correction takes the form
\begin{align}
	\delta\sigma =& \frac{e^2D}{\pi \hbar} \int_{\tau}^{\tau_\phi} \frac{B/\phi_0}{\sinh(4\pi DtB/\phi_0)} dt \notag\\
	=& \frac{e^2}{2\pi h}\left[\Psi\left(\frac{1}{2}+\frac{\tau_B}{\tau}\right) - \Psi\left(\frac{1}{2}+\frac{\tau_B}{\tau_\phi}\right)\right] 
\end{align}
where $\tau_B = \hbar/(4eDB)$ and $\Psi$ is the digamma function. The characteristic magnetic field strength needed to destroy the phase coherence of the cooperon, and thereby the WAL, corresponds to a flux quantum through the loop. Since $\tau_B \gg \tau$ the first digamma function is often replaced by its asymptotic form, resulting in the Hikami-Larkin-Nagaoka expression\cite{Hikami:1980jn}
\begin{equation}
	\delta\sigma = \frac{\alpha e^2}{\pi h} \left[ \ln\left(\frac{\tau_B}{\tau}\right) - \Psi\left(\frac{1}{2} + \frac{\tau_B}{\tau_\phi}\right)\right]. 
	\label{eq:WALB}
\end{equation}
We have introduced the factor $\alpha$ which is simply equal to $1/2$ here, but will be useful in comparing with experiments below. 

A magnetic field can also give rise to periodic oscillations in the conductance when the surface is multiply connected. This is realized for example in TI nanowires, where if the bulk is insulating the surface acts as a hollow metallic cylinder. Applying flux $\phi$ along the wire introduces a phase $2\pi n\phi/\phi_0$ to each path that loops around the cylinder $n$ times. This modifies the WAL correction in a sample of width $W$ and length $L$ to be
\begin{equation}
	\delta\sigma = \frac{e^2}{\pi h} \left[ \ln \frac{L}{\ell} + \sum_n \cos \frac{4\pi n \phi}{\phi_0} \ln \left(1-e^{-\pi n \frac{W}{L}} \right)\right],
	\label{eq:WALwire}
\end{equation}
up to an unimportant constant. The magnetoconductance oscillates with a period of $\phi_0/2$. The factor of two originates in the interference between clockwise and anti-clockwise circulating paths that each pick up a flux $\phi$. This type of oscillations where originally discussed by Altshuler, Aronov and Spivak in the context of metallic cylinders,\cite{Altshuler:1981ui} and experimentally observed by Sharvin and Sharvin.\cite{Sharvin:1981tq} For this reason they are often referred to as AAS oscillations in the literature (for a review see Ref.~\onlinecite{Aronov:1987ce}). Away from the symplectic metal limit the magnetoconductance of the TI nanowire can realize robust oscillations with period $\phi_0$, double that of the AAS oscillations. This is discussed in section~\ref{sec:AB}.

\subsubsection{Universal conductance fluctuations}
Weak anti-localization is a quantum correction to the average conductance. For a given sample, as external parameters are varied, the conductance fluctuates around the average due to quantum interference. The amplitude of the fluctuations is universal (independent of microscopic parameters) and of the order of $e^2/h$. These are the universal conductance fluctuations (UCF). Like WAL, the UCF amplitude is determined by the symmetry class and is a characteristic of the symplectic metal that is independent of topology. The UCF can be understood in terms of the path picture of the last section. We will not discuss this here and instead refer for example to Ref.~\onlinecite{Akkermans:2007vf} for details and further references.

In the context of Dirac fermions the UCF was studied theoretically in Refs.~\onlinecite{Cheianov:2007fw,Rycerz:2007fi,Kechedzhi:2008in,Kharitonov:2008gw,Rossi:2012jq}. UCF has been observed experimentally in TI's in several experiments and its temperature dependence used to estimate the phase coherence length~\onlinecite{Checkelsky:2011cx,Matsuo:2012dj,Li:2012wn}. Anomalously large conductance fluctuations were observed in large Bi$_2$Se$_3$ crystals,\cite{Checkelsky:2009de} presumably because of the large bulk conductivity. 

\subsubsection{Field theory of diffusion}
\begin{figure}[tb]
	\begin{center}
		\includegraphics[width=0.9\columnwidth]{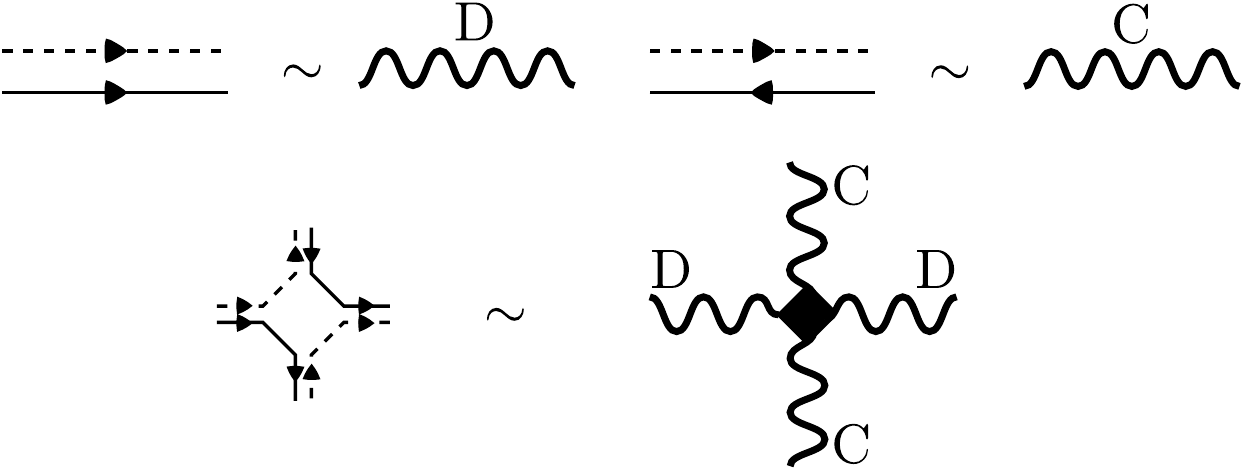}
	\end{center}
	\caption{In the field theory of diffusion, the non-linear sigma model, the propagators of the fields correspond to the diffusion and cooperon (top, cf.\ also figures~\ref{fig:wal} and~\ref{fig:diagrammatics}). The four field interaction vertex coincides with the Hikami box (bottom). Higher order interaction terms are also present (not shown).}
	\label{fig:NLsM}
\end{figure}

In our discussion of WAL we have employed a semiclassical path picture and provided its connection to diagrammatic calculations. Another theoretical approach that is commonly adapted in localization studies is the field theory approach. The field theory describing diffusion is the non-linear sigma model (NL$\sigma$M).\cite{Efetov:1997uz} The fields in the NL$\sigma$M live on a manifold that is determined by the symmetry class. In some cases, these manifolds allow the presence of a topological term, that is a term that only depends on the topology of the field configurations, in the NL$\sigma$M. The localization properties depend strongly on the presence or absence of the topological terms (cf.\ section~\ref{sec:AbsenceOfLoc}).

To have a picture in mind, we provide a connection between the NL$\sigma$M and the semiclassical path approach. The propagators of the fields in the NL$\sigma$M are given by the diffusion and the cooperon, see~\fref{fig:NLsM}. The non-linearity of the NL$\sigma$M corresponds to interactions between the cooperons and diffusions. In particular, the four field interaction vertex corresponds exactly to the Hikami box, or the crossing point. Higher order interaction vertices correspond to interference between paths with more than one self-crossing.\cite{Hikami:1981bv}

The coupling constant in the NL$\sigma$M is related to the conductivity of the system. The WAL correction, as depicted in~\fref{fig:wal}, is the one-loop renormalization of this coupling constant. Higher order correction have been calculated and in the symplectic metal the next nonzero correction is only obtained at four-loop order (the calculation has been carried out to five-loops).\cite{Hikami:1992em} This is one reason why the WAL correction~\eref{eq:WAL} describes numerical simulations very well already at not too large conductivity and system size (cf.~section~\ref{sec:AbsenceOfLoc}).

\section{Longitudinal conductivity of topological insulator surfaces}
\label{sec:cond}

In graphene, the key observations confirming the Dirac nature of the charge carriers were the anomalous quantum Hall effect and the ambipolar field effect with the accompanying minimum conductivity.\cite{CastroNeto:2009cl,DasSarma:2011br} Observation of these effects in TI's would constitute a convincing step towards verifying through transport experiments the Dirac dispersion of the surface states. Magnetic field and flux effects on quantum transport, including the quantum Hall effect, will be discussed in detail in the next two sections. In this section we focus on the zero field longitudinal conductivity and the absence of localization.

\subsection{Ambipolar field effect}
\begin{figure}[tb]
	\begin{center}
		\includegraphics[width=0.9\columnwidth]{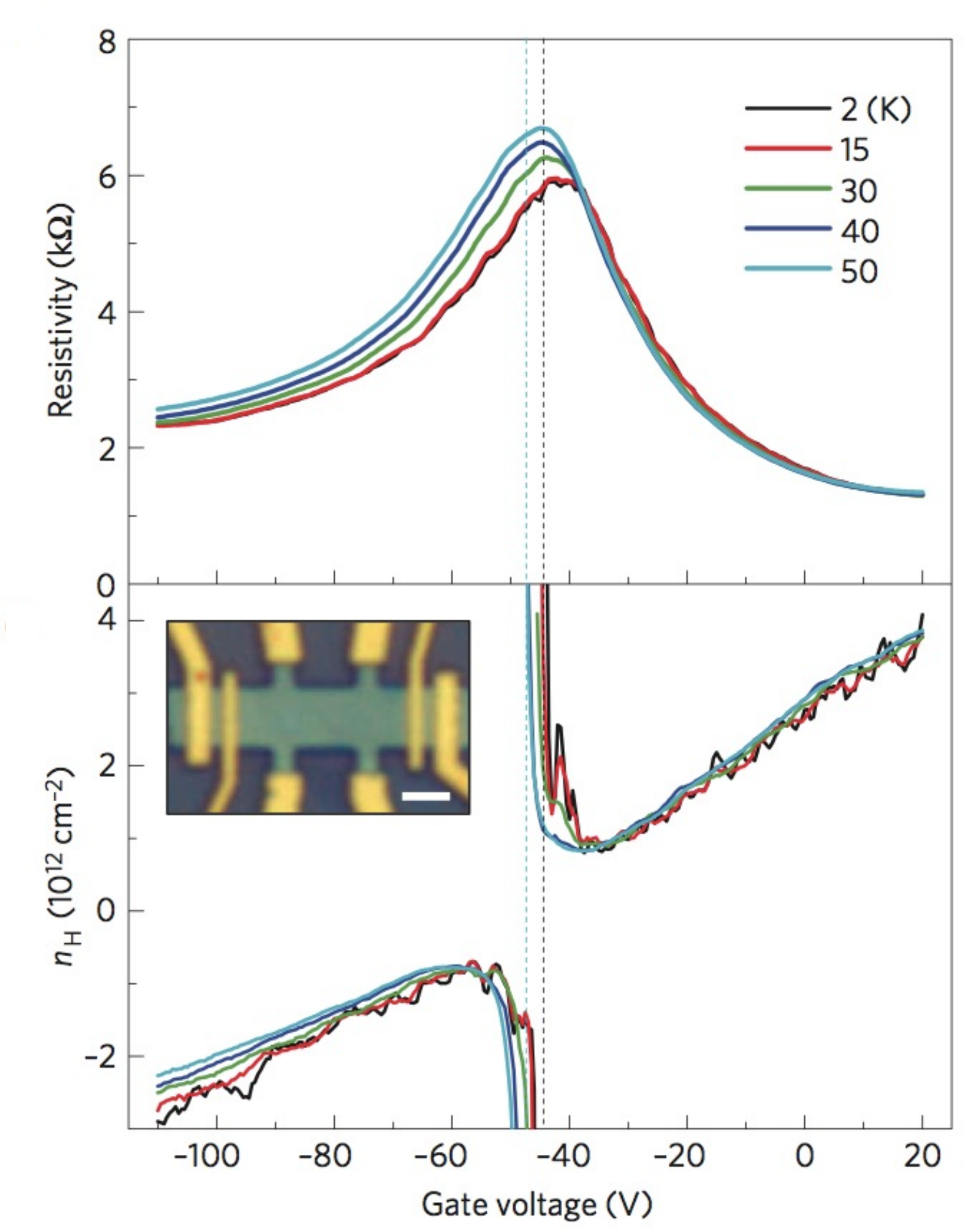}
	\end{center}
	\caption{Longitudinal resistivity of a Bi$_2$Se$_3$ thin film (about 10 nm thick) obtained by mechanical exfoliation as a function of back gate voltage at various temperatures (top). To reduce the bulk density the sample has been chemically $p$-doped with F4TCNQ molecules. The lower panel shows the carrier density as obtained by a Hall measurement. The linear dependence of the Hall density at high and low gate voltage indicates unipolar electron and hole transport. The inset shows an optical micrograph of the device with a scale bar of 2 $\mu$m. Adapted from Ref.~\onlinecite{Kim:2012eg}.}
	\label{fig:Kim12}
\end{figure}
Observation of minimum conductivity and ambipolar field effect requires the conductance to be dominated by the surface. Progress towards that end has been made by reducing the bulk density by doping and using thin films. Several groups have reported minimum conductivity and/or ambipolar field effect.\cite{Steinberg:2010kd,Ren:2011fo,Kong:2011fe,Sacepe:2011hm,Hao:2012ba,Kim:2012eg} As an example, \fref{fig:Kim12} shows experimental data demonstrating both a minimum conductivity and an ambipolar Hall field effect in Bi$_2$Se$_3$ thin films. As in graphene the minimum conductivity depends only weakly on temperature. The Hall density is linear in the gate voltage at low and high gate voltage, suggesting that the charge carriers are of only one type. Together these results indicate that the transport is through a surface state rather than being obtained from the bulk.

In graphene the minimal conductivity is understood in terms of the physics of electron and hole puddles induced by disorder (for a review see Ref.~\onlinecite{DasSarma:2011br}). The same is likely to hold true in TI transport, with nonlinearities in the spectrum away from the Dirac point playing a more important role than in graphene.\cite{Culcer:2010kf,Adam:2012dv} In fact, electron hole puddles have already been observed in STM experiments.\cite{Beidenkopf:2011jw} This is an important and current area of theoretical and experimental interest. It is, however, outside of the scope of this review which focuses more on the quantum correction to the conductivity. We will therefore not discuss it further, but rather briefly review the quantum theory of the longitudinal transport, single parameter scaling and the absence of localization. The results of these considerations will be of relevance to later discussion of WAL and quantum transport in WTI's, and is of fundamental interest by itself. 

\subsection{Absence of localization}
\label{sec:AbsenceOfLoc}
\begin{figure}[tb]
	\begin{center}
		\includegraphics[width=0.9\columnwidth]{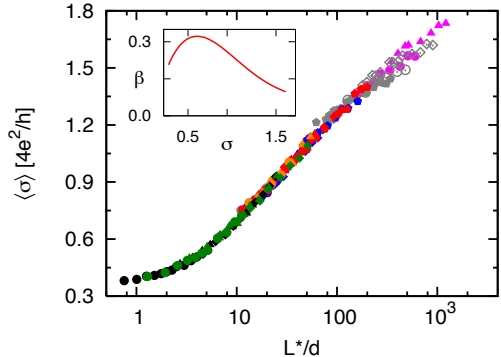}
	\end{center}
	\caption{Conductivity $\sigma$ at the Dirac point as a function of system size $L$ obtained by a numerical solution of the Dirac equation in presence of scalar disorder. The different data points are for different disorder strengths and correlation length. The system size has been rescaled by a mean free path to collapse the data on a single parameter scaling curve. At large conductivity and system size the scaling curves follows the logarithmic WAL dependence~\eref{eq:WAL} with the expected slope of $1/\pi$. The inset shows the scaling function $\beta = d\log\sigma/d\log L$ demonstrating renormalization flow to the stable symplectic metal phase. (The factor of $4$ in the conductivity scale is from the valley and spin degeneracy of graphene, and can be discarded for TI's). Adapted from Ref.~\onlinecite{Bardarson:2007iu}.}
	\label{fig:AbsenceOfLoc}
\end{figure}
The surface of a strong TI is topologically protected from Anderson localization.\cite{Fu:2007ei,Ryu:2007bu,Ostrovsky:2007hc,Bardarson:2007iu} Instead, disorder always drives the surface into the stable symplectic metal phase.\cite{Bardarson:2007iu,Ryu:2007bu} This even holds at the Dirac point, where in the clean case the density of states goes to zero and the condition for the perturbative calculation giving WAL ($k_F\ell \gg 1$) does not hold. The surface transport in TI's is therefore always diffusive at low enough temperature and characterized by WAL.

Anderson localization\cite{Evers:2008gi} generally happens in 2D electronic systems when the dimensionless conductance $g \sim 1$. In this regime, the perturbative calculations of section~\ref{sec:SymplecticMetal} are not valid and one needs to rely on numerical simulations. In~\fref{fig:AbsenceOfLoc} we show numerical results demonstrating the flow of the conductivity with increasing system size $L$ towards that of the symplectic metal.\cite{Bardarson:2007iu,Nomura:2007jb} At large enough $L$ the conductivity increases logarithmically with a slope $1/\pi$ consistent with WAL~\eref{eq:WAL}. The scaling function $\beta(\sigma) = d\ln\sigma/d\ln L$ is strictly positive,\cite{Bardarson:2007iu,Nomura:2007jb} as opposed to the normal 2D electron gas with spin-orbit coupling which has a metal-insulator transition\cite{Markos:2006gc,Evers:2008gi} (and therefore a sign change in $\beta$) at $\sigma \sim 1.4$. 

\begin{figure}[tb]
	\begin{center}
		\includegraphics[width=0.9\columnwidth]{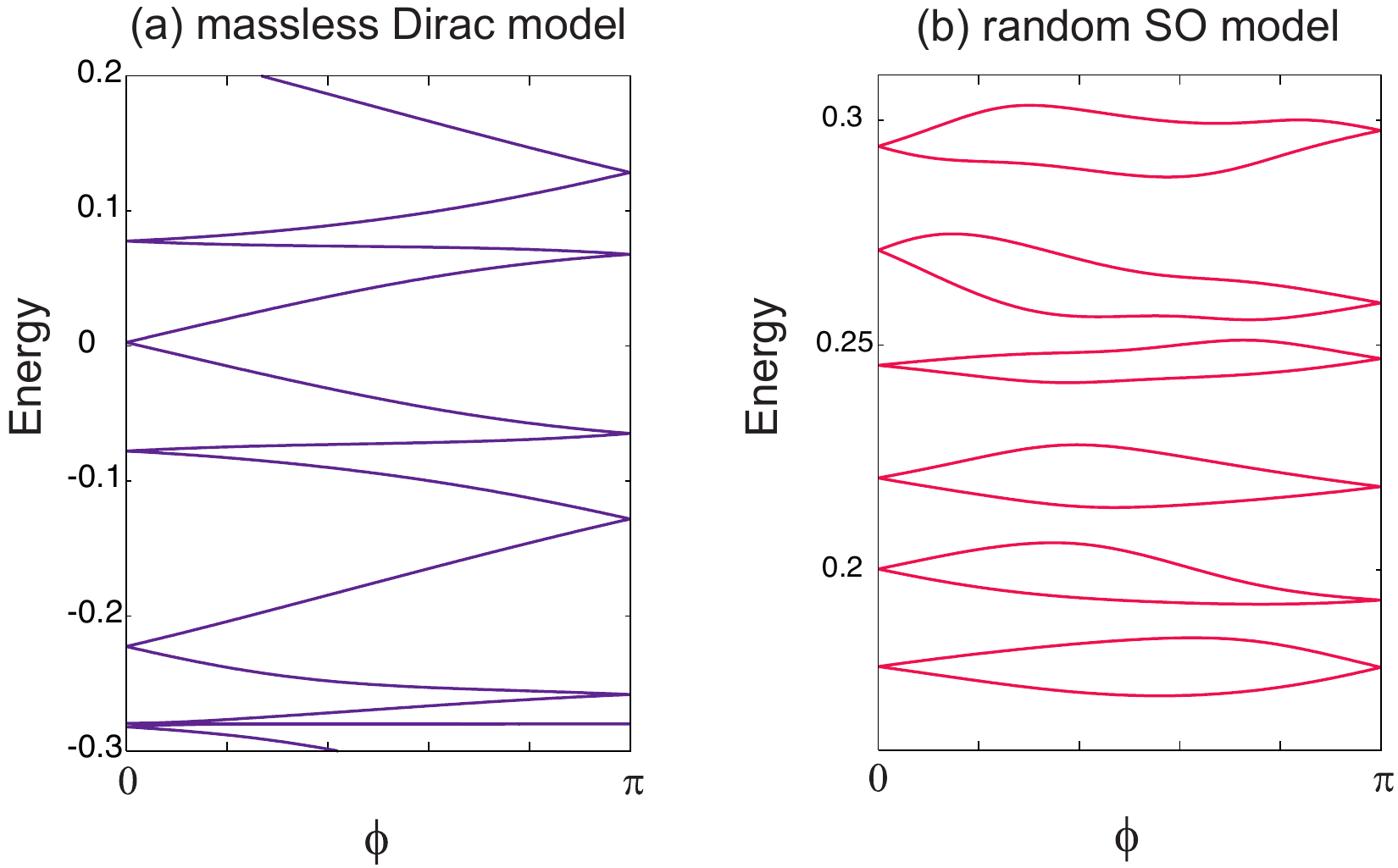}
	\end{center}
	\caption{The energy spectrum of a massless Dirac fermion (left) and a random spin-orbit coupled 2DEG (right) in presence of disorder, as a function of boundary twist angle $\varphi$ where $\psi(x,y+W) = \exp(i\varphi)\psi(x,y)$. At $\varphi = 0$ and $\varphi = \pi$ the spectrum is Kramers (doubly) degenerate. The Kramers pairs switch partners in the Dirac case but not in the 2DEG case. This is one manifestation of the $\mathbb{Z}_2$ topological classification in the symplectic class. Adapted from Ref.~\onlinecite{Nomura:2007jb}.}
	\label{fig:AbsenceOfLocEnergyLevels}
\end{figure}
The topological nature of the protection from localization can be understood by exploring the dependence of the energy spectrum on the boundary condition.\cite{Fu:2007ei,Nomura:2007jb} Specifically, \fref{fig:AbsenceOfLocEnergyLevels} shows the sensitivity of the energy levels of a disordered Dirac fermion in a finite sample to a twist in the boundary condition $\psi(x,y+W) = e^{i\varphi}\psi(x,y)$. At $\varphi =0$ and $\varphi = \pi$ the system is time-reversal symmetric and the spectrum is Kramers degenerate. Going between these two special values the Kramers pairs exchange partners.\cite{Nomura:2007jb} This pair switch is a topological property and disorder can not change this behavior. The band width $\delta_\varphi$ is always of the order of the mean level spacing $\Delta$ and by Thouless argument\cite{Lee:1985hn} the conductance is at least $g \sim \delta_\varphi/\Delta \sim 1$ and localization can not take place.  Indeed, a localized state is unaffected by changes in the boundary conditions and its energy is independent of the boundary flux $\varphi$ and $\delta_\varphi \rightarrow 0$. In a conventional 2DEG the Kramers partners do not switch pairs and localization ensues for strong enough disorder. The switch versus no switch of Kramer's partners reflects the $\mathbb{Z}_2$ nature of the topological classification of 3D time-reversal invariant TI's.\cite{Fu:2007ei}

At the field theory level, the difference between a single Dirac fermion and a 2DEG is understood to be due to the presence or absence of a topological term in the NL$\sigma$M respectively.\cite{Ryu:2007bu,Ostrovsky:2007hc} The 3D TI is thus characterized by a topological term in the effective field theory of diffusion at the surface which is in one lower dimension. Generalizing this notion to all dimensions and symmetry classes one can classify all possible topological insulators and superconductors by the possibility of realizing a topological term in the corresponding NL$\sigma$M.\cite{Schnyder:2008ez} 

In addition to the effects of generic disorder discussed here, several works, both theoretical and experimental have explored the effects of local impurities.\cite{Biswas:2010hp,BlackSchaffer:2012ca,Alpichshev:2012jk}

\section{Magnetotransport in crystals of 3D topological insulator materials}
\label{sec:field}

\subsection{Shubnikov-de Haas oscillations and quantum Hall effect}
As a prelude to discussing the Aharonov-Bohm effect and other flux effects in the following section, we now discuss standard magnetoconductance measurements on 3D topological insulators.  Electrical transport in a 3D topological insulator receives contributions from both bulk and surface states.  A goal of much current research is to find 3D topological insulator materials that are truly insulating in the bulk, as defined for example by a resistivity that shows activated behavior at low temperature (i.e., diverges exponentially as $\exp(E_g / k_B T)$, where $E_g$ is the bandgap, up to power-law factors).  This has been difficult to achieve, and in the few cases where a divergence of this form is seen, the effective bandgap is much smaller than the expected bandgap from either photoemission calculations or electric structure theory.\cite{Skinner:2012vr}

Magnetotransport measurements provide a natural means to distinguish between bulk and surface transport when the bulk retains a nonzero conductivity.  Ideally, a sample can be made sufficiently clean that both bulk and surface contributions will show magnetoconductance oscillations, the Shubnikov-de Haas (SdH) oscillations.  Such oscillations are periodic in inverse magnetic field with period
\beq
\Delta (1/B) = \frac{2 \pi e}{\hbar S_F},
\eeq
where $S_F$ is the cross-sectional Fermi surface area transverse to the applied field.  This is an area in $k$-space, which has dimensions of inverse area, so the oscillation period is essentially an area divided by the flux quantum, up to a numerical factor.  An ideal 3D topological insulator with one bulk Fermi surface and one surface Fermi surface will show two periods in the magnetoconductance oscillations, and these two oscillations will be distinguished by different areas and different angular dependences, assuming that the bulk Fermi surface is less two-dimensional than the surface one.

In a strictly two-dimensional electron system, SdH oscillations occur as a precursor of the quantum Hall effect.  On a quantum Hall plateau the diagonal resistivity $\rho_{xx}$ drops to zero (note that the tensorial nature of ${\bf \rho}$ implies that the diagonal components of both resistivity and conductivity are zero).  The locations of the Landau levels in a conventional 2DEG are at half-integer multiples of the cyclotron energy $\omega_c \propto B$, $E_n = (n+1/2) \hbar \omega_c$, with $n=0, 1, \ldots.$  At fixed Fermi level $E_F$ in the 2DEG, as is achieved if the 2DEG is in contact with a reservior, the condition for the $n$th Landau level to cross is $E_n = E_F$, or
\beq
(n+1/2) \propto \frac{1}{B}.
\eeq
So these features are equally spaced in $1/B$, but with an offset of $1/2$.  In an STI this offset is absent because the spectrum is Dirac-like: including a Zeeman term for later reference, the Landau levels for a Dirac cone of velocity $v$ occur at
\beq
|E_n| = \sqrt{(g \mu B/2)^2 + 2 |n B| \hbar e v^2},
\eeq
where the first term is the Zeeman effect and the second gives an $E \sim \sqrt{B}$ dependence.  For zero Zeeman effect, the Landau level crossings occur at
\beq
2 n B \hbar e v^2 = {E_F}^2
\eeq
so now $n \propto (1/B)$ with no offset, as promised.  Hence a ``Landau level index plot''  (``fan diagram'') can in principle distinguish between the case of a quadratic  spectrum and a Dirac spectrum.

In the SdH regime the same difference in offset shows up in the phase of the quantum oscillation.  However, in SdH oscillations the diagonal resistivity and conductivity oscillate out of phase, while in the quantum Hall regime they move in-phase.  For example, in the quantum Hall regime both are zero on a plateau as the resistivity and conductivity are purely off-diagonal tensors.  A useful reference for SdH oscillations is the book of Shoenberg;\cite{Shoenberg:1984wo} here we focus on the key differences between Dirac and ordinary (quadratic) fermions.  A basic assumption in the following conventional theory is that the only effect of the magnetic field is that its coupling to the orbital degrees of freedom is essentially as a ``probe'': it does not modify the electronic structure except through this orbital effect.  As high magnetic fields have to be used to observe oscillations, in part because the surface electron density is high and mobility is not much higher than $10^4$ V/(cm$^2$/s) in current samples even at low temperature, this assumption needs to be considered carefully, especially in possible future materials where the surface topological state results from correlation phenomena that are sensitive to magnetic field.

The two cases of SdH oscillations (normal 2DEG and Dirac fermions) can be obtained starting from the semiclassical quantization of energy levels:
\beq
S_F (E) = \frac{2 \pi e B}{\hbar} (n+\gamma)
\eeq
where $n$ is the Landau level index of the oscillation, $S_F$ is the cross-sectional Fermi surface area at energy $E$ and the shift $\gamma=1/2$ for a conventional 2DEG and $\gamma=0$ for an STI.
As emphasized by Mikitik and Sharlai,\cite{Mikitik:1999jq,Mikitik:2012ka} the shift is quantized to one value or the other depending only on the number of Dirac points; details of energetics are unimportant.  While this semiclassical formula is only generally valid for large $n$, note that for $\gamma=1/2$ in a conventional 2DEG, we obtain for Landau level $n$
\begin{align}
S_F(E) &= \pi {k_F}^2 = \frac{2 \pi e B}{\hbar} \left(n+\frac{1}{2}\right) \notag \\ &\Rightarrow E =  \hbar \omega_c \left(n+\frac{1}{2}\right)
\end{align}
consistent with the quantum Hall limit.  For the Dirac case, we obtain at Landau level $n$
\begin{align}
S_F(E) &=  \pi {k_F}^2 = \frac{2 \pi e B}{\hbar} n \notag \\ &\Rightarrow E^2 = 2 \hbar n v^2 e B,
\end{align}
again consistent with the exact calculation.

In practice the strength of the oscillations is reduced by both disorder and thermal fluctuations.  The Lifshitz-Kosevich formula can be applied to estimate the strength of the oscillations incorporating this shift; in the simplest case (for additional corrections, see Ref.~\onlinecite{Ando:1974fr}; for experimental examples, see Ref.~\onlinecite{Fang:wf} and references therein),
\beq
\Delta R \propto R_T R_D \cos\left(\frac{S_F \hbar}{eB}+2 \pi \gamma\right)
\eeq
is the oscillatory component of the resistance $R$.  Here $R_D$ is the Dingle damping factor from disorder,
\beq
R_D = e^{-\pi / \tau \omega_c},
\eeq where $\tau$ is the scattering time, and $R_T$ describes reduction due to thermal smearing.  One recent theoretical analysis\cite{Mikitik:2012ka} of SdH experiments in Bi$_2$Se$_3$ and Bi$_2$Te$_3$ specifically for this phase $\gamma$ argues that all experiments are consistent with $\gamma=0$ once effects of nonzero chemical potential and Zeeman energy are taken into account. Our discussion here has been in terms of flat infinite surfaces. The QHE effect on finite curved and multiple connected surfaces has been explored theoretically in Refs.~\onlinecite{Lee:2009do,Vafek:2011ie,Sitte:2012ib}.

There are several factors, some fundamental and some material-dependent, that complicate the observation of this physics in a 3D topological insulator.  First, while the bulk of a topological insulator probably does serve to some extent as a reservoir of electrons for the surface state at a fixed chemical potential, this ignores surface charging effects and also ignores the dependence of bulk electron properties on applied magnetic field.  Second, in transport the overall measured conductivity or resistivity always include a significant bulk contribution in current materials, although impressive progress has been made in reducing the bulk contribution.

Another complication is that the $g$ factor may be significantly larger than 2 (as large as 30-50); this is known to be the case for bulk electrons in Bi$_2$Te$_3$, Bi$_2$Se$_3$, and HgTe materials as a consequence of the same relativistic effects that lead to the TI behavior.  Electronic structure calculation of the $g$ factor of the surface state is technically challenging, but STM experiments and most transport experiments are consistent with $g \leq 10$ so that the Zeeman effect is relatively weak compared to the orbital effects.  A ``smoking gun'' for the Zeeman effect is that the zeroth Landau level's energy is solely determined by the Zeeman effect, so observation of the zeroth level at high fields would allow a fairly direct measurement of the Zeeman effect, but this requires starting with a sample whose chemical potential is close to the Dirac point.  Some intrinsic interest of the Zeeman effect in the surface state is that it does not commute with the starting Hamiltonian, unlike in either graphene or a conventional 2DEG (assuming spin-orbit coupling can be neglected).  As a result, the Zeeman effect modifies the eigenstates, not just their energies, and opens up an energy gap even if the orbital field is excluded.

\subsection{Experimental observations of Shubnikov-de Haas oscillations and the quantum Hall effect}
Early observations of quantum oscillations were dominated by the bulk.\cite{Analytis:2010jl,Eto:2010ca,Butch:2010fu} From the SdH oscillations the anisotropic nature of the bulk Fermi surface is verified. This anisotropy also shows up in angular-dependent magnetoresistance oscillations.\cite{Eto:2010ca,Taskin:2010kv} For these samples with large bulk conductance, a good characterization of these bulk contributions is required to extract any possible surface contribution. This may require very large fields.\cite{Analytis:2010kl}

\begin{figure}[tb]
	\begin{center}
		\includegraphics[width=0.8\columnwidth]{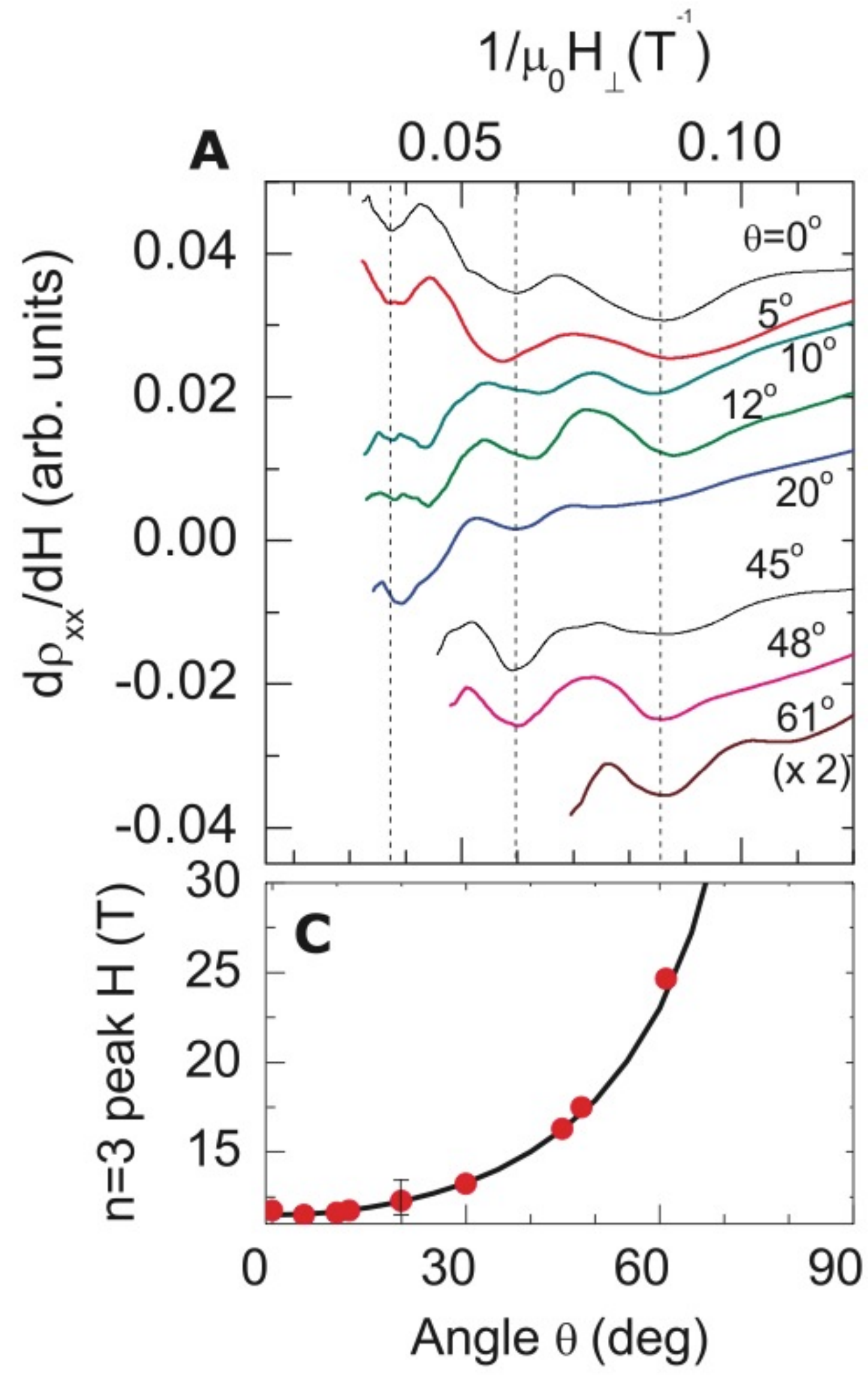}
	\end{center}
	\caption{The resistivity derivative $d\rho_{xx}/dH$ as a function of the perpendicular component of the magnetic field $1/H_\perp$ (upper panel) where $H_\perp = H\cos\theta$ and $\theta$ is the tilt angle between the magnetic field direction and the surface normal. The various curves correspond to different tilt angles. Non metallic samples of Bi$_2$Te$_3$ are obtained by growing with a weak compositional gradient. The lower panel shows the tilt angle dependence of the position of the third Landau level, which is consistent with the $1/\cos\theta$ expected from a 2D Fermi surface (solid line). Adapted from Ref.~\onlinecite{Qu:2010ixa}.}
	\label{fig:Qu10}
\end{figure}
\begin{figure}[tb]
	\begin{center}
		\includegraphics[width=0.99\columnwidth]{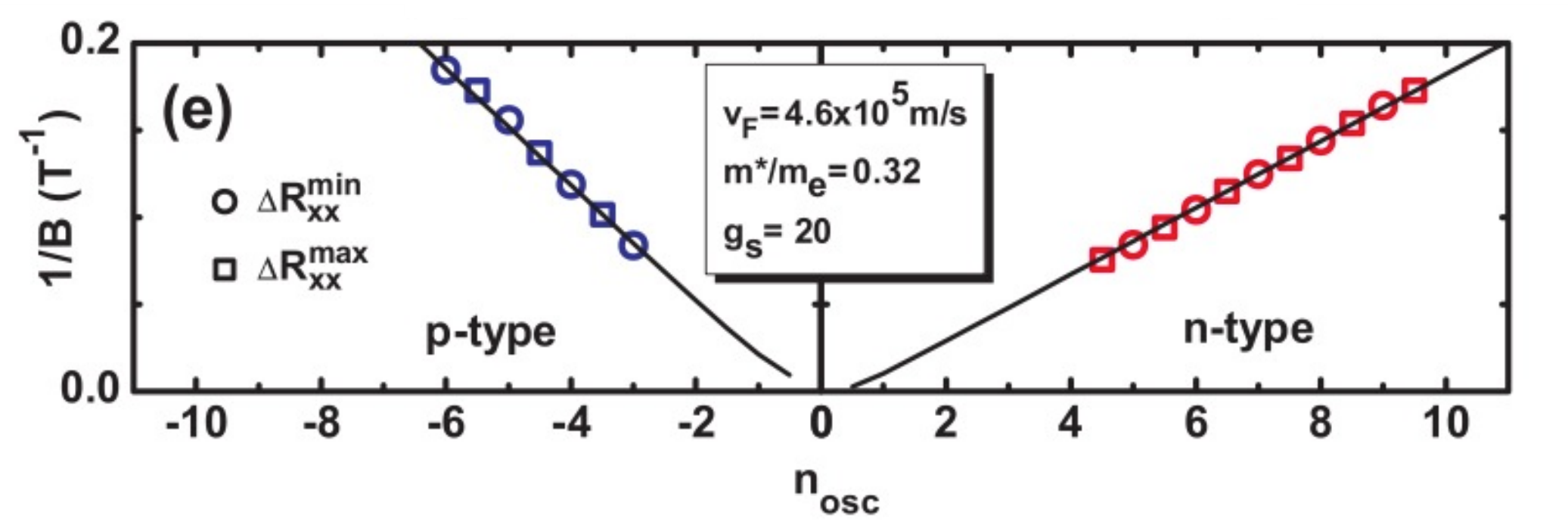}
	\end{center}
	\caption{Position of maxima and minmia of SdH oscillations (Landau level fan diagram) in a 30 $\mu$m thick Bi$_{1.5}$Sb$_{0.5}$Te$_{1.7}$Se$_{1.3}$ sample. The blue (left) data is obtained 5 hours after cleaving when the surface chemical potential is such that the surface is $p$-type. After 720 hours the chemical potential has shifted and the surface is now of $n$-type as demonstrated by the red (right) data. The solid lines are fit to theory of SdH taking into account nonlinearity of the spectrum and is consistent with the expected Berry phase of $\pi$. Adapted from Ref.~\onlinecite{Taskin:2011dx}.}
	\label{fig:Taskin11}
\end{figure}
Alternatively, in samples with larger surface contribution, clear 2D SdH oscillations are observed.\cite{Qu:2010ixa,Analytis:2010kl,Ren:2010ji,Taskin:2011dx,Ren:2011fo,Wang:2012dg,He:2012cy,Xiong:2012ht,Xiong:2012fc,Petrushevsky:2012ji} The 2D nature of the SdH oscillations is for example revealed by tilting the magnetic field. The 2D SdH only depends on the perpendicular component of the magnetic field, and therefore the position of resistance maxima varies as $1/\cos\theta$. An example of SdH oscillations in Bi$_2$Te$_3$ is shown in~\fref{fig:Qu10}. The position of the maxima in the resistance can in principle be used to extract the value of the Berry phase through the shift $\gamma$ introduced in the last section. Most experiments have been interpreted to be consistent with a non-trivial Berry phase, though an accurate extraction of its value is complicated by nonlinearites of the spectrum and a nonzero $g$ factor.\cite{Taskin:2011gw,Xiong:2012ht,Petrushevsky:2012ji} An example of a Landau level fan diagram is given in~\fref{fig:Taskin11}. In many samples the doping level of the surface is time dependent due to aging effects. By using this time dependence it was possible to probe both electron and hole transport in the same sample.\cite{Taskin:2011dx}  

\begin{figure}[tb]
	\begin{center}
		\includegraphics[width=0.9\columnwidth]{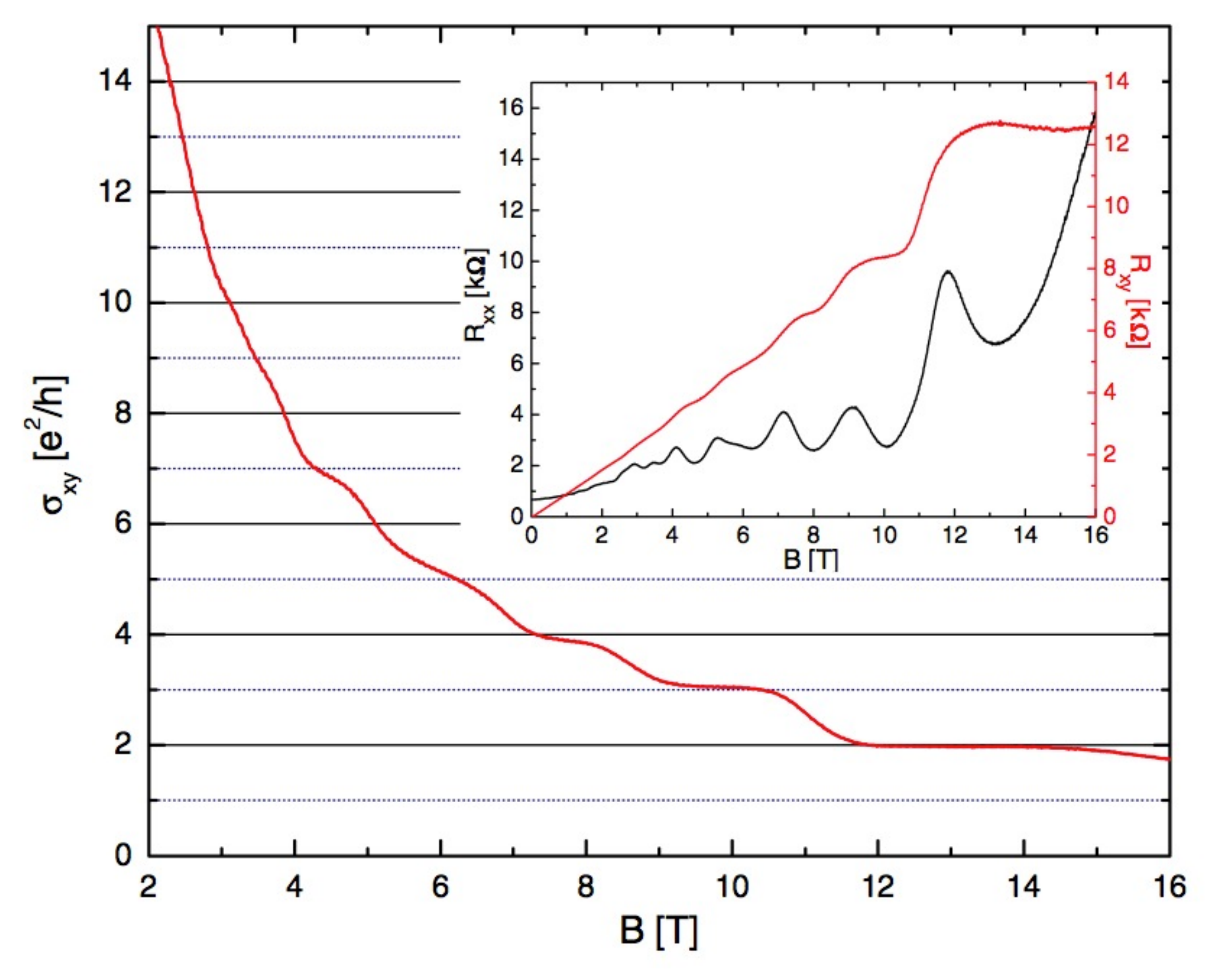}
	\end{center}
	\caption{Hall conductivity as a function of magnetic field for a strained 70 mm thick HgTe sample, measured at 50 mK. The strain opens up a gap in the nominally semimetallic 3D HgTe, which is estimated theoretically to be of the order of 20 meV. The inset shows the Hall and longitudinal resistances. While the Hall conductivity has some well developed plateus the longutidinal resistance does not go all the way to zero. Adapted from Ref.~\onlinecite{Brune:2011hi}.}
	\label{fig:Brune11}
\end{figure}
While there are several works that report SdH oscillations, a fully developed QHE is rare though signatures of the QHE have been reported.\cite{Cheng:2010ie,Hanaguri:2010fm,Brune:2011hi} In particular, \fref{fig:Brune11} shows the Hall conductance obtained in strained HgTe with plateaus at multiples of $e^2/h$. The appearance of plateaus at even values of $\sigma_{xy}/(e^2/h)$ is interpreted to be a consequence of a different filling factor in the top and bottom surfaces. The longitudinal conductance does not go to zero at the plateaus, presumably due to the side surfaces remaining conductive. 

A quantized Hall effect and SdH oscillations with $1/\cos\theta$ dependence have also been observed in highly doped samples, known to be dominated by the bulk,\cite{Cao:2012kc} and interpreted as evidence of layered bulk transport. As opposed to the surface data, these layered SdH oscillations where consistent with zero Berry phase. Finally we mention that related quantum oscillations have also been observed in other quantities, such as magnetization.\cite{Taskin:2009ev,Kim:2011eo,Lawson:2012wr}

\section{Magnetic flux effects on quantum transport in topological insulator surfaces}
\label{sec:flux}
In this section we discuss two related effects of magnetic flux on quantum transport, weak anti-localization and Aharonov-Bohm or Altshuler-Aronov-Spivak oscillations. While these are often viewed as separate effects, they all arise from the AB phase of the wave function and quantum interference between paths. In particular, WAL and AAS oscillations are exactly the same phenomena except they are realized in samples with different topology (flat versus multiply connected). When coupled with the Berry phase, AB oscillations of seemingly different nature can arise as discussed below. The distinction between the AB and the AAS oscillations is really that the former is realized when the particle motion is ballistic or pseudo-diffusive, while the latter are obtained in the diffusive state. An important difference between the two cases is that in the diffusive state, as we discussed in the introduction, one can not distinguish between a TI and a normal 2DEG. This is in principle possible with the AB oscillations. 

\subsection{Weak anti-localization in thin films}
Since the surface state is always driven into the symplectic metal phase, one expects to observe weak anti-localization. WAL can both be observed as a negative magnetoconductance~\eref{eq:WALB} and as a logarithmic temperature enhancement of the conductivity~\eref{eq:WAL}. We will discuss the two approaches separately in the following. To avoid having the WAL signal being completely masked by bulk effects most experiments to date work with thin films. These are mostly Bi$_2$Se$_3$ films\cite{Hirahara:2010da,Chen:2010iz,Liu:2011kv,Checkelsky:2011cx,Chen:2011cm,Wang:2011ji,Kim:2011fe,Steinberg:wb,Liu:2012eq,Matsuo:2012dj,Takagaki:2012hv,Zhang:2012vq,Bansal:2011us} but studies with Bi$_2$Te$_3$\cite{He:2011ix} and Bi$_2$(Se$_x$Te$_{1-x}$)$_3$\cite{Cha:2012jt} have also been reported. To further reduce the bulk conductivity the films are sometimes additionally doped, for example with Ca,\cite{Checkelsky:2011cx} Pd\cite{Wang:2011ji} or Cu,\cite{Takagaki:2012hv} or the Fermi level is moved into the bulk gap with the help of a gate voltage on a back gate\cite{Chen:2010iz,Checkelsky:2011cx,Chen:2011cm} or a top gate.\cite{Steinberg:wb} The effect of magnetic doping and the resulting crossover to weak localization (WL) has also been studied, both experimentally\cite{Liu:2012eq} and theoretically.\cite{Lu:2011iz}

\subsubsection{Magnetic field dependence of weak anti-localization}
\begin{figure}[tb]
	\begin{center}
		\includegraphics[width=0.99\columnwidth]{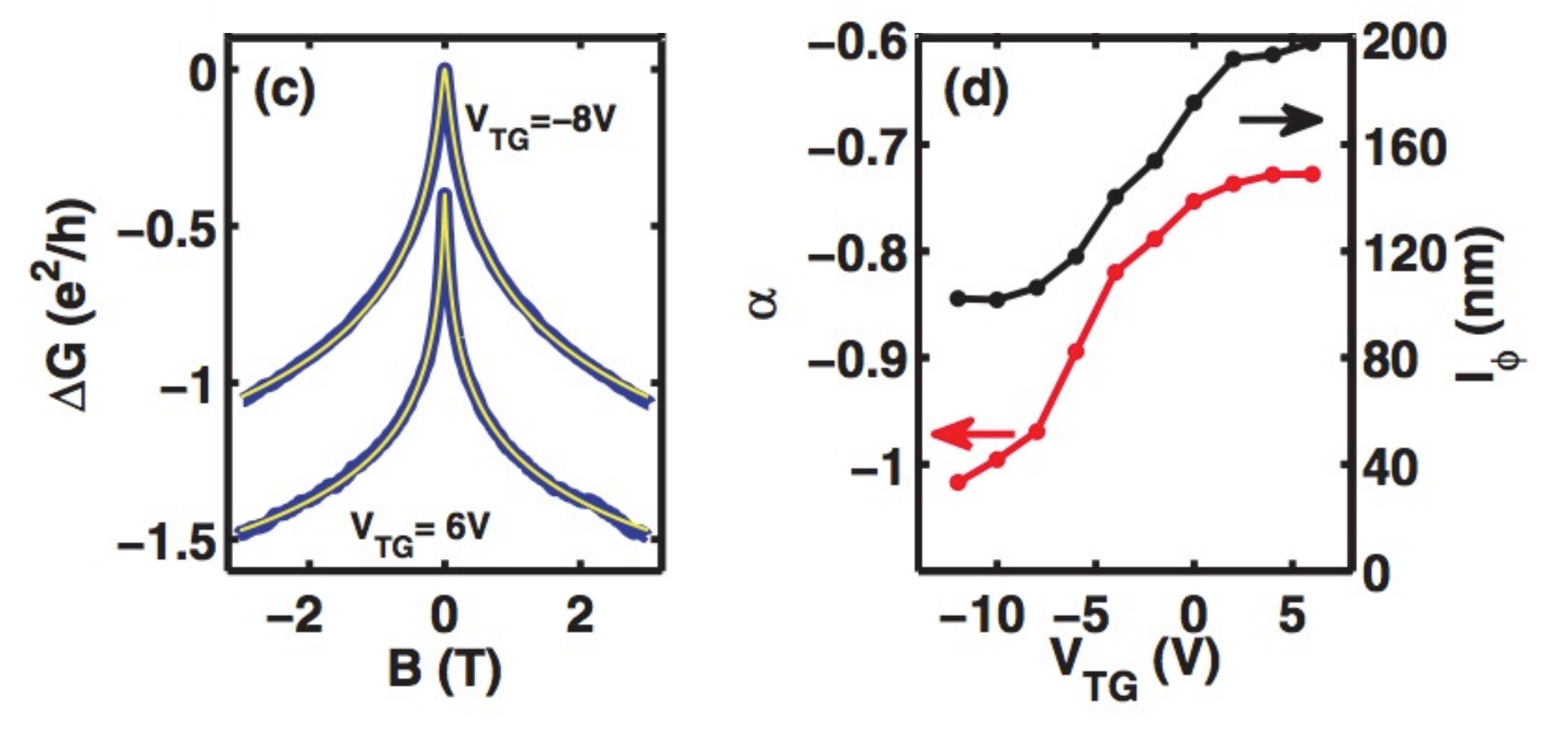}
	\end{center}
	\caption{Magnetoconductance of a Bi$_2$Se$_3$ thin film grown on a Si(111) substrate, for two different top gate voltages (left). The data has been fitted to the WAL expression~\eref{eq:WALB} to obtain the amplitude $\alpha$ and the phase coherence length $\ell_\phi$. (The definition of $\alpha$ in this plot is of opposite sign to the one adopted in this review). The plot on the right shows the extracted values of $\alpha$ and $\ell_\phi$ as a function of the top gate voltage. A large negative gate voltage induces a depletion layer between the top surface and the bulk channel which then independently contribute to the WAL. Adapted from Ref.~\onlinecite{Steinberg:wb}.}
	\label{fig:WALexp}
\end{figure}
Most experiments observe a negative magnetoconductance consistent with~\eref{eq:WALB} but with a prefactor that varies roughly in the range $0.3 < \alpha < 1.2$.\cite{Hirahara:2010da,Chen:2010iz,Liu:2011kv,Checkelsky:2011cx,Chen:2011cm,Wang:2011ji,Kim:2011fe,Steinberg:wb,Liu:2012eq,Matsuo:2012dj,Takagaki:2012hv,Zhang:2012vq,He:2011ix,Cha:2012jt} The value of $\alpha$ can be tuned by as much as a factor of two in a single sample by varying gate voltage\cite{Chen:2010iz,Chen:2011cm,Steinberg:wb} (see~\fref{fig:WALexp}). Both the broad range of values obtained for $\alpha$ and its gate voltage tunability can be understood by carefully taking into account the bulk contribution and the coupling between surface and bulk. Some of this physics was discussed theoretically in Ref.~\onlinecite{Lu:2011ja} and later in detail in Ref.~\onlinecite{Garate:2012fm}. We give a simplified discussion of the results of Ref.~\onlinecite{Garate:2012fm} here and refer to the original works for further details.

When the thickness of the film $t \ll \ell_\phi$, the bulk is effectively two dimensional and contributes 2D quantum correction of its own. Generally, due to the strong spin-orbit coupling in TI materials, this is also a WAL of the form~\eref{eq:WALB} with $\alpha = 1/2$. However, if the bulk is only weakly doped and intervalley and spin scattering in the bulk is negligible, a WL with $\alpha = -1$ is obtained instead. Roughly speaking, this is because the spin of the eigenstates close to the band edge depends only weakly on the momentum. The Berry phase from the spin rotation is thus absent. It is likely that in most current samples the bulk is in a parameter range where WAL with $\alpha = 1/2$ is expected. Strong coupling with the surface state will also generally give rise to WAL, even if one expects WL from the isolated bulk. In the following we assume the bulk is in the WAL regime. We note though that WL has been observed in ultrathin films (4 - 5 quintuple layers) in Bi$_2$Se$_3$ films\cite{Zhang:2012vq} and (Bi$_x$Sb$_{1-x}$)$_2$Te$_3$ films.\cite{Lang:UHUFrjDZ} It does not seem likely that these observations are explained by the bulk contribution. In particular, in the former study the presence of WL is substrate dependent, suggesting that details of the hybridization of the top and bottom surface and their environment plays an important role.

If the surface and bulk are completely decoupled their contribution to WAL will simply add and a magnetoconductance~\eref{eq:WALB} with $\alpha = 1$ is expected. In contrast, if the bulk and surface are strongly coupled, $\tau_{\rm sb} \ll \tau_\phi$ with $\tau_{\rm sb}^{-1}$ the surface to bulk tunneling rate, they act as a single channel with $\alpha = 1/2$. With intermediate surface-bulk coupling a value of $\alpha$ somewhere between the two limiting values is expected.  Top and bottom surface can also contribute independently and have different coupling with the bulk and different phase coherence lengths. This explains the observed range of values for $\alpha$ obtained in the experiments. The gate voltage dependence of $\alpha$ is likely a consequence of a variation of the surface-bulk coupling. The gate voltage induces a depletion region below the surface, reducing the tunnel coupling of the surface to the bulk.\cite{Steinberg:wb,Garate:2012fm}

Our discussion in this section (and most analysis of current experiments) has been simplistic in the sense that it assumes that the observed WAL correction to the conductance can still be described by~\eref{eq:WALB}, just with a renormalized value of $\alpha$. While there are limits in which this is justified,\cite{Garate:2012fm} the bulk, top and bottom surface generally have different phase coherence lengths and the WAL is described rather by a sum of terms of the form~\eref{eq:WALB}. Nevertheless, the good qualitative agreement between theory and experiment can be taken as a convincing evidence for surface transport. However, since WAL is a characteristic of the symplectic metal that is independent of the topological properties of the system realizing the phase, a different type of experiments is needed to probe the topological properties of TI surface states through transport.

\subsubsection{Temperature dependence of weak anti-localization and the effect of electron-electron interactions}
In contrast to the good agreement between magnetoconductance experiments and the theory of the symplectic metal, the temperature dependence of the conductance does not show WAL following~\eref{eq:WAL}. Instead a WL behavior, that is~\eref{eq:WAL} with $\alpha$ negative, is observed.\cite{Liu:2011kv,Chen:2011cm,Wang:2011ji,Takagaki:2012hv} This may possibly be explained by taking electron-electron interactions into account, which we have so far ignored in our discussion. Electron-electron interactions also give a logarithmic correction to the conductance, the so called Altshuler-Aronov (AA) correction.\cite{Altshuler:1985ta} This correction is relatively independent of a magnetic field and can therefore in principle be differentiated from the WAL correction by the application of a magnetic field, as has been observed experimentally.\cite{Chen:2011cm}

These experimental observations are a least qualitatively consistent with the AA scenario, but convincing quantitative comparison is lacking. Other explanations for these observations have not been ruled out, or even explored theoretically.\cite{Pal:2012if} If, however, these are really signatures of electron-electron interactions, it opens up the possibility of observing interesting correlation physics. For example, it has been suggested that the combination of the localization effect of interactions and the topological protection of the surface will drive the surface into a stable critical state, characterized by a universal value of conductivity.\cite{Ostrovsky:2010eq}

\subsection{Aharonov-Bohm effect in topological insulator nanowires}
\label{sec:AB}
An ideal TI nanowire, with an insulating bulk, can be thought of as a hollow metallic cylinder. As explained in the introduction, a magnetic flux along the length of the wire leads to periodic oscillations in the conductance as a function of flux with period $h/2e$: the AAS oscillations. These oscillations are simply another manifestation of WAL as they are a consequence of quantum interference between clockwise and anti-clockwise circulating paths. As such, the AAS oscillations are characteristic of the symplectic metal and can not directly probe the topological properties of the surface state. Observation of these oscillations is however a strong indication of surface transport. 

The AAS oscillations are expected when the transport is diffusive. Oscillations with period $h/e$, twice that of the AAS oscillations, can be realized when transport is close to being ballistic, or when the Fermi level is at or close to the Dirac point. The basic physics behind these latter oscillations, to be discussed in the next section, is the Berry phase and the perfectly transmitted mode. Observation of these oscillation would constitute an indirect measurement of the surface state Dirac fermion's Berry phase.  

\subsubsection{Berry phase and the perfectly transmitted mode}
A TI nanowire differs from a conventional quantum wire in that it can host an odd number of transmission modes and therefore a perfectly transmitted mode with conductance of $e^2/h$. Due to the Berry phase the perfectly transmitted mode is realized at a flux through the wire equal to a half integral number of flux quanta.\cite{Ran:2008dv,Rosenberg:2010dj,Ostrovsky:2010eq} If all other modes in the wire give a negligible contribution to the conductance the conductance will oscillate with a period of $h/e$ with a minimum at flux $\phi = 0$.\cite{Bardarson:2010jl,Zhang:2010bd}

\begin{figure}[tb]
	\begin{center}
		\includegraphics[width=0.9\columnwidth]{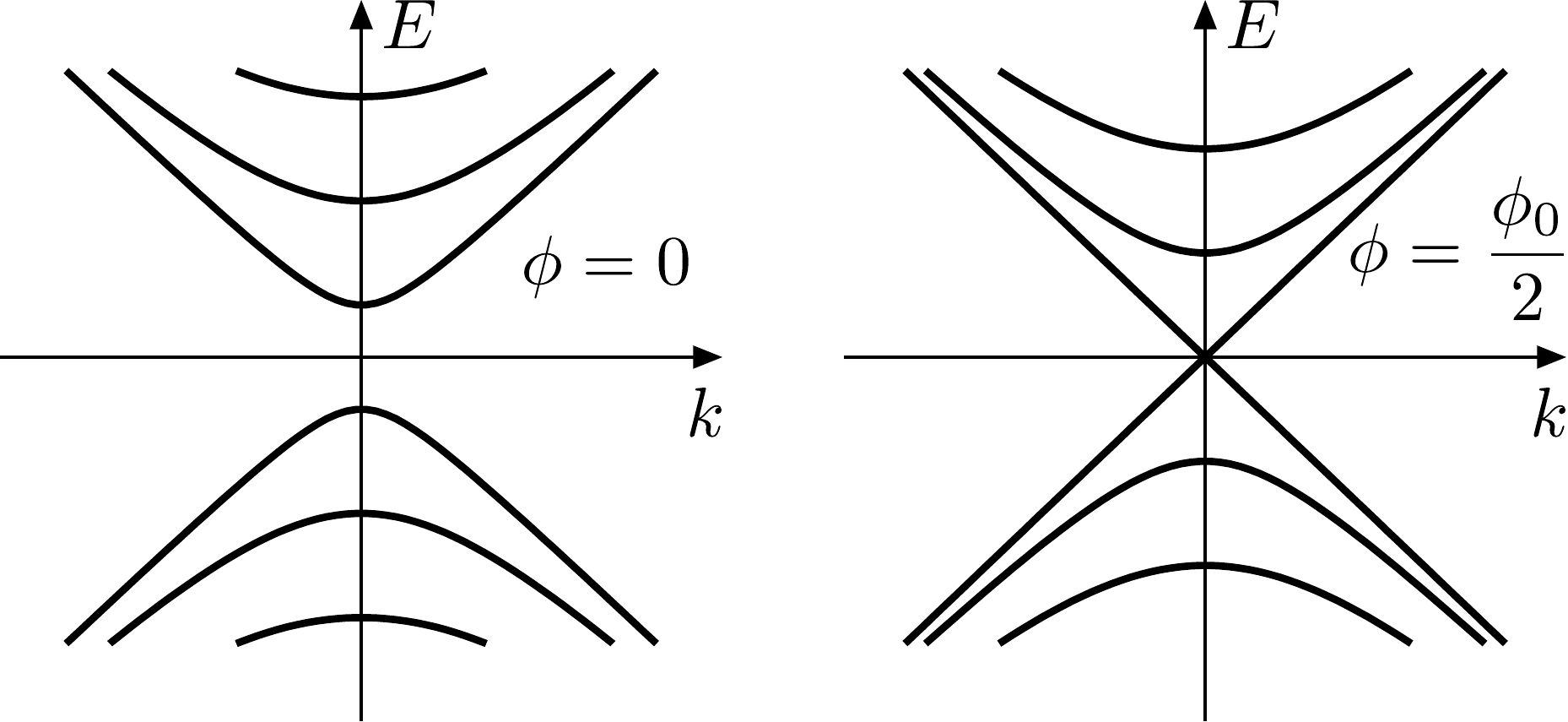}
	\end{center}
	\caption{A schematic drawing of the energy spectrum of a TI nanowire for two different values of flux along the wire. The horizontal axis $k$ is the momentum along the length of the wire. The different branches of the spectrum correspond to different values of the transverse momentum $q_n = 2\pi n/W$ with $W$ the circumference. All the lines are therefore doubly degenerate, corresponding to clockwise and anti-clockwise circulation around the wire, expect for the linear curves which correspond to $q = 0$ and are nondegenerate. Consequently the number of transmission modes at flux $\phi = \phi_0/2$ is odd but even ad $\phi = 0$. Adapted from Ref.~\onlinecite{Bardarson:2010jl}.}
	\label{fig:Bardarson10a}
\end{figure}
To demonstrate this in an explicit model we consider the effective theory of the surface state (some of the results below have also been obtained with a 3D lattice model\cite{Zhang:2010bd}), described by a single Dirac cone
\begin{equation}
	H = v \mathbf{p}\cdot\mathbf{\sigma}.
	\label{eq:HDirac}
\end{equation}
In addition to the kinetic term a curvature induced spin connection term, that describes how the spin rotates as it moves along the surface, is generally present.\cite{Lee:2009do} For a cylindrical surface this term can be completely absorbed into the boundary condition
\begin{equation}
	\psi(x,y+W) = e^{i\pi}\psi(x,y),
	\label{eq:BCap}
\end{equation}
which is now anti-periodic. Here $x$ is the coordinate along the wire and parallel to the flux, $y$ is the transverse coordinate and $W$ is the circumference. The anti-periodicity is a consequence of the fact that the spin lies in the tangent plane to the surface and therefore rotates by an angle of $2\pi$ when looping around the surface. Because of the antiperiodic boundary condition the spectrum is gapped as schematically shown in~\fref{fig:Bardarson10a}. Within the same approximation, the effect of applying a flux along the wire is simply to change the boundary condition by the Aharanov-Bohm phase
\begin{equation}
	\psi(x,y+W) = e^{i(2\pi\phi/\phi_0 + \pi)}\psi(x,y).
	\label{eq:BCapflux}
\end{equation}
A flux of half a flux quantum, $\phi = \phi_0/2$, cancels the Berry phase and the spectrum becomes gapless (\fref{fig:Bardarson10a}). Crucially, the number of modes is now odd and there is necessarily at least one perfectly transmitted mode with conductance $e^2/h$.
\begin{figure}[tb]
	\begin{center}
		\includegraphics[width=0.9\columnwidth]{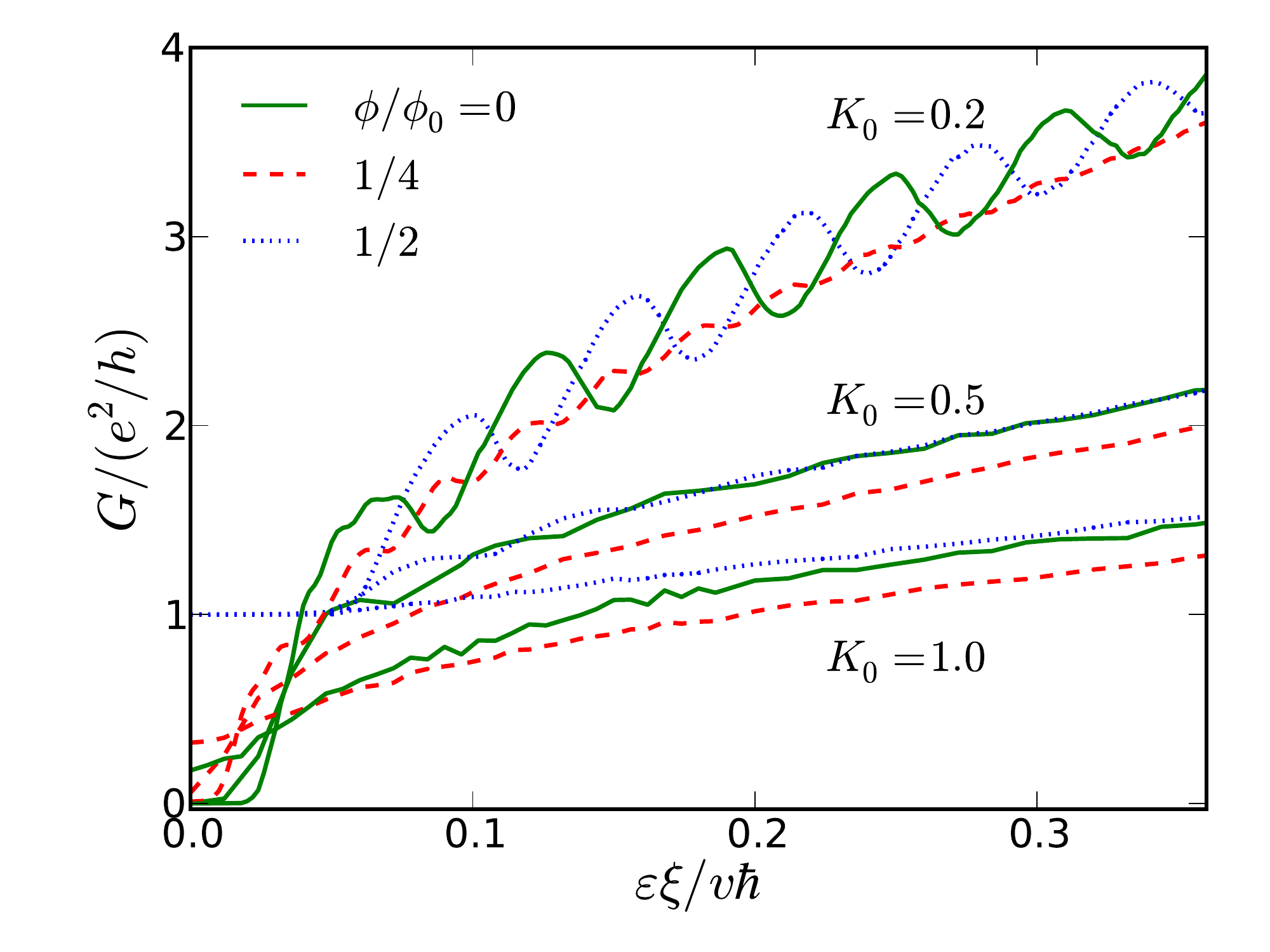}
	\end{center}
	\caption{Two terminal conductance of a TI nanowire as a function of Fermi energy $\epsilon = \hbar v k_F$, for various disorder strengths $K_0$. For each value of $K_0$ the conductance is plotted for three different values of flux $\phi$. When the conductance at $\phi = 0$ and $\phi = \phi_0/2$ is the same the magnetoconductance oscillates with a period of $h/2e$ (AAS oscillations). This happens at moderate to large disorder strength and at doping away from the Dirac point. Magnetoscillations with period $h/e$ are otherwise obtained. At the Dirac point the conductance always has a maximum at flux $\phi = \phi_0/2$ but away from the Dirac point a maximum at either $\phi = 0$ or $\phi = \phi_0/2$ is obtained depending on the doping level. The data is obtained by numerically solving the Dirac equation in presence of disorder for a wire with $W = 100\xi$ and $L = 200\xi$ where $\xi$ is the correlation length of the disorder. The oscillations as a function of Fermi energy away from the Dirac point at $K_0 = 0.2$ have a period which corresponds to the mean level spacing $\Delta = \hbar v/W$. Adapted from Ref.~\onlinecite{Bardarson:2010jl}.}
	\label{fig:AASandAB}
\end{figure}

\Fref{fig:AASandAB} shows the conductance as a function of doping as obtained by a numerical simulation.\cite{Bardarson:2010jl} The simulation includes a Gaussian disorder potential $V(r)$ that satisfies $\langle V(r)V(r^\prime) \rangle = K_0 (\hbar v)^2 \exp[(r-r^\prime)^2/(2\xi^2)]/2\pi\xi^2$. $K_0$ is a dimensionless measure of the disorder strength and $\xi$ is the length scale characterizing the disorder (typical size of electron-hole puddles). For each value of $K_0$ the conductance is given at three values of the flux, which determine the period of possible oscillations. Three different regimes are observed: i) If disorder is large enough such that transport is diffusive the AAS oscillations dominate. The conductance is the same at $\phi = 0$ and $\phi = \phi_0/2$. ii) At the Dirac point conductance is dominated by the perfectly transmitted mode. The conductance oscillates with a period of $h/e$ and has a minimum at zero flux. iii) At weak disorder and large enough doping, such that the conductance is no longer dominated by the perfectly transmitted mode, the conductance oscillates with a period of $h/e$. In this regime the conductance can have either a maximum or a minimum at zero flux, depending on the doping level. The conductance oscillates as a function of doping with a period equal to the level spacing $\Delta = \hbar v 2\pi/W$. These oscillations are smeared by temperature when $k_BT \approx \Delta$ and the conductance at $\phi = 0$ and $\phi = \phi_0$ is equal. The conductance then oscillates with a period of $h/2e$.

\subsubsection{Experimental observations of Aharonov-Bohm oscillations}
\begin{figure}[tb]
	\begin{center}
		\includegraphics[width=0.9\columnwidth]{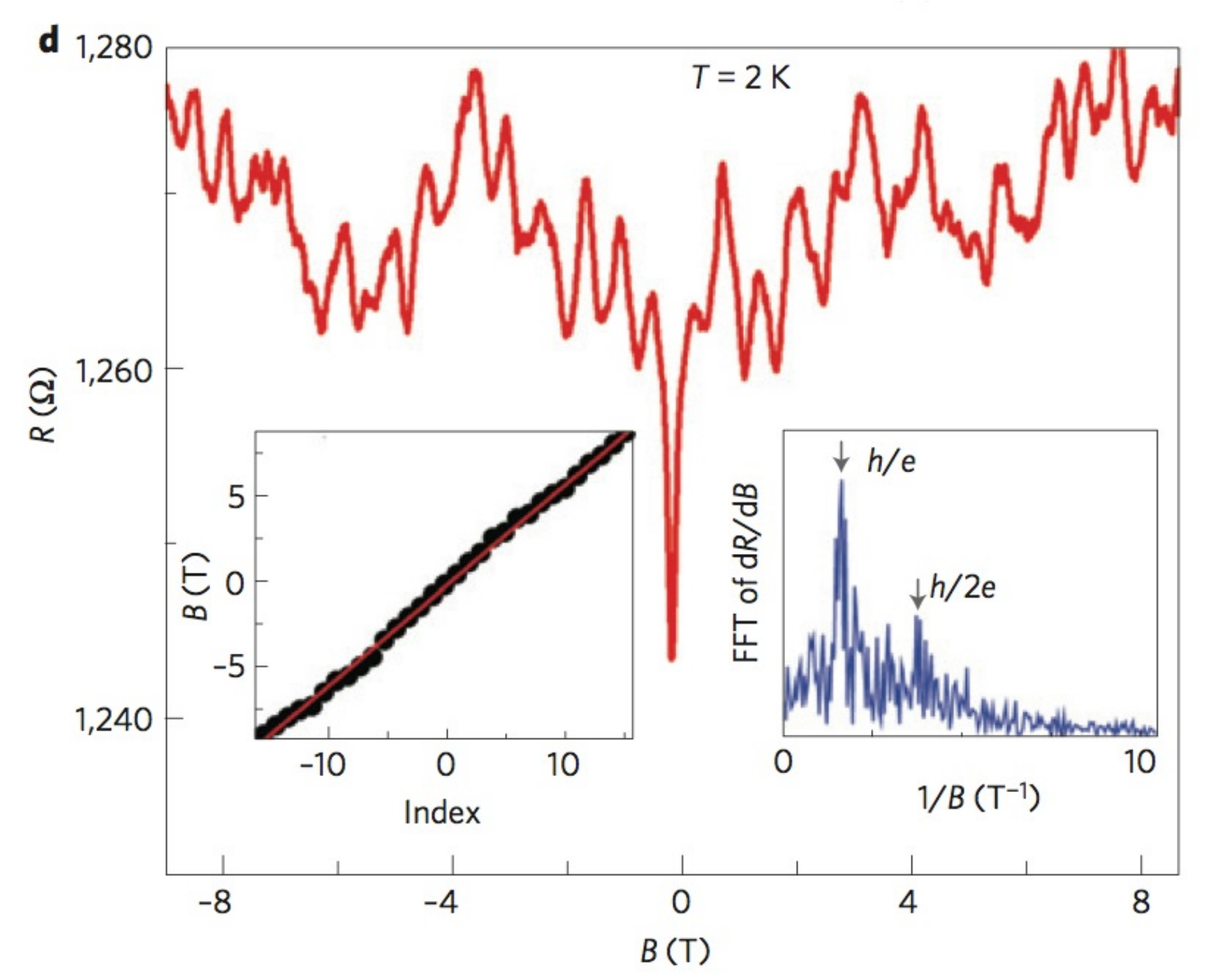}
	\end{center}
	\caption{Aharonov-Bohm oscillations in the resistance as a function of magnetic field of a Bi$_2$Se$_3$ nanoribbon of a rectangular cross section (120 nm $\times$ 50 nm) and length $\sim$ 2 $\mu$m. The left inset shows the position of resistance minima and the right inset the Fourier transform of the derivative $dR/dB$, with a strong peak at a magnetic field corresponding to a flux of $h/e$ and a weaker peak at flux $h/2e$. The resistance has a minimum at zero flux and the relatively sharp increase of resistance around zero flux is indicative of WAL, possibly from the bulk which density is nonzero. Adapted from Ref.~\onlinecite{Peng:2009jm}.}
	\label{fig:Peng2011}
\end{figure}
Aharonov-Bohm oscillations have been experimentally observed in nanowires (ribbons, plates) of Bi$_2$Se$_3$,\cite{Peng:2009jm} Bi$_2$Te$_3$,\cite{Xiu:2011hq,Li:2012jk} and $\beta$-Ag$_2$Te.\cite{Sulaev:2012wa} Related periodic oscillations have also been observed in arrays of ringlike structures.\cite{Qu:2011dn} In the wires, the magnetoconductance shows periodic oscillations superimposed on the UCF (see \fref{fig:Peng2011} for example data). Fourier transform reveals a strong $h/e$ period and a weaker $h/2e$ period.

All the samples are doped away from the Dirac point which in Bi$_2$Te$_3$ is buried in the valence band. In Ref.~\onlinecite{Xiu:2011hq} a back gate was used to tune the Fermi level and reduce the bulk contribution to the conductance. The AB oscillations become more pronounced as the surface contribution is enhanced. If these observations are to be consistent with the theory of the last section, the samples would need to be in the weak disorder limit. As a necessary condition the mean level spacing must be large compared to the temperature broadening. From the circumference of the samples one estimates $\Delta \sim 10K - 70K$ while the measurement temperature is usually $T \sim 2K$. In Ref.~\onlinecite{Li:2012jk} a crossover from $h/e$ to $h/2e$ oscillations is observed at $T \approx 5K$ for a sample with $\Delta \approx 10K$.

In the weak disorder regime the conductance at zero flux will be either maximum or minimum depending on the sample. In Refs.~\onlinecite{Peng:2009jm,Xiu:2011hq,Sulaev:2012wa} the conductance has a maximum while in Ref.~\onlinecite{Li:2012jk} it has a minimum. A change in the sign of the magnetoconductance with varying gate voltage was not observed,\cite{Xiu:2011hq} though a systematic exploration of this effect was not undertaken. A clear ``smoking gun'' signature of Dirac fermion transport in this regime would be the observation of the gate voltage oscillations accompanied by a $\pi$ shift in the magnetoconductance. 

Taken together these results are consistent with the theory of doped and weakly disordered TI wires. A more systematic exploration of doping and disorder strength dependence of the AB oscillations is needed to firmly confirm this picture. In particular, the robust $h/e$ oscillations at the Dirac point and the accompanying indirect measurement of the Berry phase is still to be observed.

\section{Related problems and future directions}
\label{sec:related}
We end this review by briefly exploring a couple of related topics. These are quantum transport in WTI and the Josephson effect in TI surfaces.

\subsection{Transport at the surface of a weak topological insulator}
\label{sec:WTI}
So far we have mainly focused on the transport properties of strong topological insulators (denoted in this section with STI to clearly differentiate them from the WTI). STI's have an odd number of Dirac cones at their surface, while weak TI's have an even number of Dirac cones. A WTI may be realized e.g.\ in layered semiconductors, such as KHgSb.\cite{Yan:2012vo} Topological crystalline insulators which surface electronic structure is also described by an even number of Dirac cones, have recently been proposed\cite{Fu:2011ia,Hsieh:2012tq} and observed.\cite{Dziawa:tv,Xu:tj,Tanaka:wj} In addition a thin film of STI can be considered to be a WTI. 

A WTI can be thought of as being obtained by stacking 2D QSHE layers. The helical edges states of the QSHE couple together to form the surface state of the WTI. By construction this surface state will not be found on every face of the sample. If the number of layers is odd the number of propagating modes in the surface is also odd and the surface hosts a perfectly transmitted mode. The conductance is at least $e^2/h$ and the surface can not localize. The energy levels have a similar switching behavior as in STI.\cite{Ringel:2012hz} Instead of localization the surface of a WTI flows, under rather general conditions to be discussed below, into the symplectic metal.\cite{Mong:2012du} In this sense the WTI behaves very much like an STI. The effect of interactions on WTI surface states was studied in Ref.~\onlinecite{Liu:2012jg}.

The surface state of a WTI can be modeled by two Dirac cones
\begin{equation}
	H = v \tau_0 \otimes \mathbf{\sigma} \cdot \mathbf{p} + V(\mathbf{r}).
	\label{eq:HWTI}
\end{equation}
Here $\mathbf{p}$ is the momentum operator and the Pauli matrices $\sigma$ and $\tau$ act in spin and valley space respectively. The disorder potential $V$ can be written as
\begin{equation}
	V = \sum_{\alpha\beta} V_{\alpha\beta}(\mathbf{r}) \tau_\alpha \otimes \sigma_\beta.
	\label{eq:VWTI}
\end{equation}
Of all the possible terms, only $\openone\otimes\openone, \tau_x\otimes\openone, \tau_z\otimes\openone$ and $\tau_y\otimes\sigma_\gamma$ with $\gamma = x,y,z$ do not break the time-reversal symmetry which is given by $\mathcal{T} = \openone\otimes i\sigma_y\mathcal{C}$ with $\mathcal{C}$ the complex conjugate, and satisfies $\mathcal{T}^2 = -1$. Of these, only $\tau_y\otimes\sigma_z$ opens up a gap in the energy spectrum and we therefore refer to the average of this term as the mass $m \equiv \langle \tau_y\otimes\sigma_z \rangle$. The disorder terms are independently distributed with $\langle \delta V_{\alpha\beta}(\mathbf{r})\delta V_{\alpha\beta}(\mathbf{r}^\prime)\rangle = g K(\mathbf{r}-\mathbf{r}^\prime)$ where $\int d\mathbf{r} K(\mathbf{r}) = 1$. 

\begin{figure}[tb]
	\begin{center}
		\includegraphics[width=0.9\columnwidth]{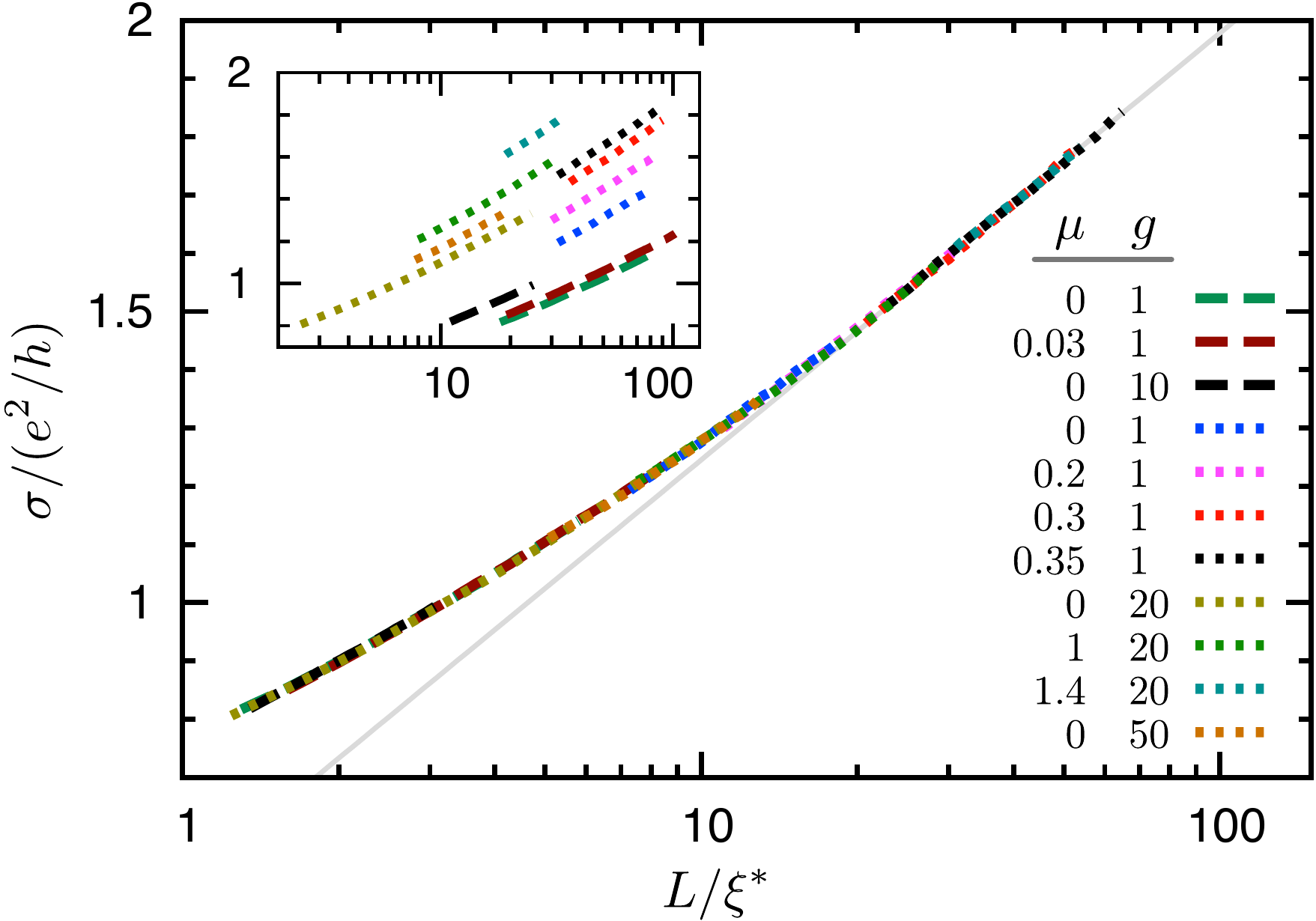}
	\end{center}
	\caption{Conductance as a function of scaled system size for various chemical potentials $\mu$ and disorder strengths $g$ at zero mass $m = 0$. The inset shows the raw data before scaling. The data is obtained by a numerical simulation of the Dirac equation~\eref{eq:HWTI}. At large system size the conductivity approaches the WAL form~\eref{eq:WAL} with slope $1/\pi$ (grey solid line). The conductivity follows single parameter scaling with a strictly positive beta function with an absence of Anderson localization, just as in the strong topological insulator. Adapted from Ref.~\onlinecite{Mong:2012du}.} 
	\label{fig:Mong12}
\end{figure}

For a random potential the mass is zero. A nonzero mass requires the surface potential to be commensurate with an even number of unit cells, and is therefore necessarily zero for an odd number of layers. In this way the mass of the 2D theory connects with the even-odd argument above. Indeed, in the absence of mass disorder drives the surface state into the symplectic metal following a single parameter scaling.\cite{Mong:2012du} \Fref{fig:Mong12} shows the result of a numerical simulation of~\eref{eq:HWTI} for various disorder strengths and doping levels including the Dirac point. The system size, horizontal axis, has been scaled by the mean free path to reveal a collapse of the raw data (inset) onto the scaling curve. At large system sizes the conductivity follows the logarithmic dependence of the WAL~\eref{eq:WAL} with a slope of $1/\pi$, just as in an STI. This is expected since the two Dirac cones are strongly mixed by the disorder and  was verified with an explicit diagrammatic calculation in Ref.~\onlinecite{Ringel:2012hz}. Note that while the WTI is similar to graphene in the sense that both systems have two Dirac cones, they differ in the sign of the square of the time-reversal operator. For a WTI $\mathcal{T}^2 = -1$ but for graphene $\mathcal{T} = 1$. Intervalley coupling in graphene therefore leads to weak localization\cite{McCann:2006ip} and eventually localization.\cite{Aleiner:2006fk}

If $m$ is nonzero, the surface can be localized by disorder. A metal insulator transition is obtained at a conductivity $\sigma \approx 1.4$, just as in a regular spin-orbit coupled 2DEG.\cite{Markos:2006gc,Evers:2008gi} The conductance distribution at the transition is also the same as in the 2DEG.\cite{Evers:2008gi} This is consistent with the two systems being described by the same effective field theory of diffusion, the symplectic NL$\sigma$M in the absence of a topological term.\cite{Ryu:2012ej} The fact that localization is avoided at $m = 0$ suggest that the full scaling flow may be described by a two parameter scaling, as in the quantum Hall effect\cite{Huckestein:1995gl}. This possibility has been explored in Ref.~\onlinecite{Mong:2012du}. If the disorder becomes strong enough that nonlinear in momentum terms need to be added to the Dirac theory~\eref{eq:HWTI} localization is expected.

The NL$\sigma$M of a WTI does not have the topological term that protects the STI from localization.\cite{Ryu:2012ej} However, it has recently been argued that for localization to happen, a proliferation of vortex like topological defects in the field configurations is necessary.\cite{Fu:2012wb} Such vortex proliferation is forbidden by average discrete symmetries which are equivalent to having a zero mass, consistent with the observed absence of localization. 

\subsection{Superconductivity and topological insulators}

Another exciting direction that merits a review article of its own is how the proximity effect from a conventional $s$-wave superconductor has unusual effects on topological insulators and related electronic structures such as strongly spin-orbit coupled semiconductors in a magnetic field.  Here we mention only some aspects of this problem that relate to the discussion above of the Aharonov-Bohm effect.  In a non-superconducting nanowire made from a topological insulator, there is a protected mode when one half of a flux quantum pierces the nanowire.\cite{Rosenberg:2010dj,Ostrovsky:2010eq}  The same metallic mode is there in the inverse problem where space is filled by a 3D topological insulator that has a cylindrical hole in it, pierced by one half of a flux quantum.  What happens to this system when the 3D topological insulator is driven superconducting?  One half of a flux quantum is of course the magnetic flux associated with an elementary vortex in a charge-$2e$ superconductor, so the setup is quite natural.

The answer is that the metallic mode becomes gapped along the length of the vortex, but there are zero-energy Majorana excitations at the two ends of the vortex\cite{Cook:2011ij,Hosur:2011iv,Chiu:2011dt,Cook:2012wj} (i.e., where it leaves the topological insulator).  Their emergence in this system can be understood in two ways.  Looking at the surface state of the topological insulator, the proximity effect converts the single-sheet Fermi surface into a ``topological superconductor'' with one Majorana fermion excitation in each vortex core.\cite{Sato:2003hz,Fu:2008gu}  An ordinary spinless fermion excitation can be broken up into two Majorana fermions, which are their own Hermitian conjugates.  This topological superconducting layer is like a time-reversal symmetric version of the 2D ``p+ip'' superconductor known to support Majorana fermions in vortex cores.\cite{Read:2000iq}  An alternate picture of where the Majorana fermions come from is as tail states of the gapped vortex core.\cite{Kitaev:2007gb} For recent reviews on realizations of Majorana modes in condensed matter systems, see Refs.~\onlinecite{Alicea:2012wg,Beenakker:tp}.  

As a first step in the direction towards the observation of the Majorana mode, and of intrinsic interest by itself, proximity effect and a Josephson current through TI surfaces has been demonstrated.\cite{Zhang:2011bl,Veldhorst:2012db,Veldhorst:2012cq,Williams:2012bv,Yang:2012eo,Yang:2012ua} Both Shapiro steps in the presence of microwave irradiation\cite{Veldhorst:2012cq} and Fraunhofer interference pattern induced by a magnetic field\cite{Veldhorst:2012db,Williams:2012bv} have been observed. The temperature dependence of the data in Ref.~\onlinecite{Veldhorst:2012cq} was more consistent with ballistic transport than diffusive transport, and the supercurrent was found to be mainly carried by the surface state, even if the normal state transport was heavily influenced by the bulk states. This servers as an additional motivation for Josephson physics as a probe of the unique properties of the normal state physics which is the focus of this article. In addition, not all aspects of the observed data are fully understood\cite{Williams:2012bv} which should spur on further theoretical and experimental efforts in this line of work.

\section{Summary}
In this report we have reviewed recent theoretical and experimental progress towards the verification of Dirac fermion surface states in topological insulators, through various magnetotransport effects. We have adapted a positive outlook on the status of the field. While there is certainly a significant number of transport experiments that show no reliable signs of surface transport for various reasons, the evidence for Dirac fermion transport and the accompanying Berry phase is mounting. Material improvements are rapid and the first steps in combining TI's with superconductors are being taken. TI transport therfore has the promise to be fertile grounds for exploring and realizing interesting physics.  

At the theoretical level, many of the transport properties of TI surfaces result from disorder driving it into the symplectic metal phase. The symplectic metal is characterized for example by weak anti-localization. However, since this phase is also realized in traditional two dimensional electron gas in the presence of spin-orbit coupling, the nontrivial Berry phase of the Dirac fermion is not easily realized in this limit. The Berry phase is revealed in principle in quantum oscillations such as Shubnikov-de Haas oscillations and, in TI nanowires, Aharonov-Bohm oscillations. We have discussed these effects in detail in the main text.

Signatures of surface transport in large bulk crystals are weak if not absent. This is due to significant bulk conduction, coming from unavoidable bulk doping. Instead, the most convincing data is obtained in thin films, where the surface can have a significant contribution to the total conductivity. Observation of weak anti-localization fitting that expected from the symplectic metal, and 2D SdH oscillations strongly suggest surface transport, though one should keep in mind that the bulk of thin films is often effectively 2D also. Comparison between theory, taking into account the bulk, and experiments suggests the presence of at least two transport channels, bulk and surface. Aharonov-Bohm oscillations have been observed in TI nanowires. These oscillations can in principle reveal the presence of the Berry phase, but this has not been attained with current data. SdH oscillations are in most cases consistent with a nontrivial Berry phase, though the data can be difficult to interpret due to various correction. Finally, the experimentally observed temperature dependence of WAL is seemingly inconsistent with the symplectic metal, possibly pointing towards important effects of electron-electron interactions.

In this review we have focused on the surface theory and mostly ignored the bulk. This reflects our view that eventually insulating bulks will be realized, and that the surface transport contains the more fundamental physics. Lastly, we mention that in addition to the reviews already cited in the introduction,\cite{Hasan:2010ku,Moore:2010ig,Qi:2011hb,Hasan:2011hs} a couple of focused reviews have appeared recently\cite{Culcer:2012ib,Tkachov:2012tw} which discuss transport of TI's from a different perspective.

\section*{Acknowledgments}
We gratefully acknowledge all our collaborators on the subject of transport in topological insulators, in particular C.~W.~J.~Beenakker, P.~W. Brouwer, and R.~S.~K.~Mong. The authors were supported by the LBNL Thermoelectrics Program of DOE BES and the DARPA TI (UCLA) program (JHB) and by NSF DMR-1206515 (JEM). 

\bibliography{refs}

\end{document}